\documentclass[twocolumn,showpacs,amsfonts,aps,prc,floatfix,nofootinbib,superscriptaddress,groupedaddress]{revtex4-1}
\usepackage{bbm}
\usepackage{float}  
\usepackage{array}
\usepackage{booktabs}
\usepackage{makecell}
\usepackage[colorlinks,linkcolor=blue,citecolor=blue]{hyperref}
\usepackage{amsmath}
\usepackage{mathrsfs}
\usepackage{amssymb}
\usepackage{graphicx,epsfig,latexsym,overpic,amssymb,color}
\usepackage{multirow}
\usepackage[section]{placeins}
\hypersetup{
	colorlinks=true,
	urlcolor=blue,
}
\usepackage[caption=false]{subfig} 

\preprint{\today}
\begin{document}
\title{Probing high-order deformation effects in neutron-deficient nuclei $^{246,248}$No with improved potential-energy-surface calculations}

\author{Jin-Liang Guo}
\affiliation{School of Physics, Zhengzhou
University, Zhengzhou 450001, China}
\author{Hua-Lei
Wang}\email{wanghualei@zzu.edu.cn} \affiliation{School of Physics, Zhengzhou University, Zhengzhou 450001, 
China}
\author{Kui Xiao}
\affiliation{School of Physics, Zhengzhou
University, Zhengzhou 450001, China}
\author{Zhen-Zhen Zhang}
\affiliation{School of Physics, Zhengzhou
University, Zhengzhou 450001, China}
\author{Min-Liang Liu}
\affiliation{Key Laboratory of High Precision Nuclear Spectroscopy,
Institute of Modern Physics, Chinese Academy of Sciences, Lanzhou
730000, China} \affiliation{School of Nuclear Science and
Technology, University of Chinese Academy of Sciences, Beijing
100049, China}

\begin{abstract}
\label{abstract}
The high-order deformation effects in even-even $^{246,248}$No are investigated by means of pairing
self-consistent Woods-Saxon-Strutinsky calculations using the potential-energy-surface (PES) approach in an extended deformation space $(\beta_2, \beta_3,\beta_4,\beta_5,\beta_6,\beta_7, \beta_8)$. Based on the calculated two-dimensional-projected energy maps and different potential-energy curves, we find that the highly even-order deformations have an important impact on both the fission trajectory and energy minima, while the odd-order deformations, accompanying the even-order ones, primarily affect the fission path beyond the second barrier. Relative to the light actinide nuclei, nuclear ground state changes to the superdeformed configuration but the normally-deformed minimum, as the low-energy shape isomer, may still be primarily responsible for enhancing nuclear stability and ensuring experimental accessibility in $^{246,248}$No. Our present investigation indicates the nonnegligible impact of high-order deformation effects along the fission valley and will be helpful for deepening the understandings of different deformation effects and deformation couplings in nuclei, especially in this neutron-deficient heavy-mass region.

\end{abstract}
%
\maketitle
%
\section{Introduction}
\label{Introduction}

Determining the limits of nuclear stability and expanding the map of known isotopes is one of the main goals of modern nuclear physics~\cite{Wang_2021,Zhang2024b, Liu_2025, Wan2025}. The single-particle structure is of great importance for nuclear stability, in particular for heavy nuclei. As it is known, the superheavy nuclei (with the atomic number $Z \geq 104$) exist only due to quantum shell effects originating from the non-uniform distribution of single-particle levels. Further, single-particle energies depend sensitively on nuclear shape (equivalently, nuclear mean field), which is usually parameterized by a set of deformation parameters, so that it is rather necessary to treat the deformations as accurately as possible, especially for a heavy nuclear system, in theoretical description.

Indeed, the mechanism of spontaneous symmetry breaking allows to represent nuclei as non-spherical shapes. Numerous experiments have indicated that nuclei can possess not only axially or nonaxially quadrople deformations but also nonaxially or axially octupole deformations and hexadecapole deformations ~\cite{Butler1996,Wang2012a,Li2023,Song2023,Wu2024,Schenke2024,Li2024,Jia2024,Wei2024}. Nuclear spectra, moments, and electromagnetic matrix elements are usually used for verifying such deformation properties~\cite{Butler1996,Stephens1955,Wang2012a}. The importance of high-order deformation, e.g., the hexacontetrapole deformation $\beta_{6}$, has been revealed in describing the ground states~\cite{Patyk1991,Moller2016,Wang2022} and the excited states including both the multi-quiparticle high-K states~\citep{Liu2011,He2019} and collective rotational states~\cite{He2019,Liu2012,Zhang2013,Xu2024}. For instance, Xu et al~\cite{Xu2024} recently probed the importance of the coupling between the high-order deformation $\beta_6$ and odd-order deformation $\beta_3$ in rotating $^{252,254}$No. In the spontaneous fission process, the effect of higher multipolarity shape parameters in nuclei with $100 \le Z \le 114$ has also been revealed~\citep{Lojewski1999}.

So far, the new generation of experimental facilities have served for many years to explore the limits of stability of high proton-number ($Z$) and/or high isospin ($T$) nuclei, including the measurements of theire stucture properties. Theoretically, the main methods include macroscopic-microscopic (MM) models and microscopic theory, e.g, cf. Refs.~\cite{Moller1995,Bender2003,Wang2014}. Prior to this work, we have performed some PES and total-Routhian-surface (TRS) calculations in multidemensional deformation spaces, e.g., ($\beta_2, \gamma, \beta_4$), ($\beta_2, \beta_3, \beta_4, \beta_5$) and some exotic deformation spaces~\cite{Wang2012a,Li2023,Song2023,Wei2024,Chai2018a,Yang2016,Chai2018b,Yang2022}. In the present paper, using an improved PES calculation with the inclusion of higher multipolarity deformations, we will focus on investigating the $^{246,248}$No nuclei, locating at both heavy and drip-line actinide region. The $^{248}$No nucleus is the most neutron-deficient even-even No isotope which has already been synthesized experimentally~\cite{NNDC} but the half-life and structure properties remain unknown, and its neighboring even-even $^{246}$No nucleus is expected to be synthesized as the next candidate. In Ref.~\cite{Lopez-Martens2022}, it was reported that a large reduction of more than five and six orders of magnitude of the ground-state fission half-lives was found between $^{252}$No and $^{254}$No and, following this trend, the fission half-life of the ground state of the more neutron-deficient even-even No isotopes will be extremely short, making it experimentally inaccessible and very close to the 10$^{-14}$-s limit of existence of an atom. However, a recent study within a cluster model pointed out that the neutron-deficient $^{247,248}$No nuclei are relatively stabe with respect to spontaneous fission and no abrupt decreases in their fission half-lives~\cite{Rogov2024}.   

Concerning the development of the PES approach, our primary contribution in this work is to extend the deformation space ($\beta_2, \beta_3, \beta_4, \beta_5$) to further cover high-order $\beta_6,, \beta_7$ and $\beta_8$ degrees of freedom, mainly including the calculation modifications of new Hamiltonian matrix, surface and Coulomb energies of nuclear liquid drop. The rest of this paper is organized as follows. The theoretical framework is described in Sec.~II. The calculated results and related discussion are presented in Sec.~III. Finally, Sec.~IV summarizes the main conclusions of present project.

\section{Theoretical method}
The general procedures of PES calculations (even with rotation, e.g., TRS) within the framework of MM models are standard and have been summarized in e.g., Refs~\cite{Xu98,Meng2022a,Meng2022b,Zhang2024a,Su2024}. In what follows we present briefly the realization of the PES approach, focusing on the leading lines and some basic definitions.

Firstly, let us briefly review one of widely-used techniques of nuclear shape (potential) parameterization. Namely, one can define the nuclear surface in terms of the spherical-harmonic basis expansion as, 
\begin{equation}
    \Sigma:
    R(\theta,\phi)
    =
    R_0c({\alpha})
    \Big[
    1
    +
    \sum_{\lambda}
    \sum_{\mu=-\lambda}^{+\lambda}
    \alpha_{\lambda\mu}
    Y_{\lambda\mu}(\theta,\phi)
    \Big],
                                                                  \label{eqn.01}
\end{equation}
where the expansion coefficients $\alpha_{\lambda\mu}$ are usually called ``deformation parameters” (also, ``deformations” in short). The ensemble of all the adopted deformation parameters $\{\alpha_{\lambda\mu}\}$ is usually abbreviated to $\alpha$. The radius parameter
$R_0 = r_0 A^{1/3}$ (here, $r_0 = 1.2$ fm), gives an approximation of the effective nuclear spherical radius in Fermi, and the auxiliary function $c(\alpha)$ assures the conservation of the nuclear volume, e.g., the volume enclosed by the nuclear surface $\Sigma$ is equal to the volume of the corresponding spherical nucleus (independent of the actual shape). To avoid the possible confusions, it may be worth noting that, similar to the coordinate space, a vector can be projected onto the axes; in the deformation space expanded by spherical harmonics, the total deformation $\beta$ and deformation $\beta_\lambda$ at $\lambda$ order are usually defined by $\beta =\sqrt{\sum_\lambda \beta^2_{\lambda} }$ and $\beta_\lambda=\sqrt{\sum_\mu \alpha^2_{\lambda\mu} }$, respectively~\cite{Myers66,Ryssens2023}. For the axially symmetric shape, e.g., considered in this project, the deformation $\beta_\lambda$ equals $\alpha_{\lambda 0}$ due to $\alpha_{\lambda\mu \neq 0} = 0 $. In this work, we consider the deformation degrees of freedom $\beta_{6,7,8}$ and spherical harmonics $Y_{\lambda_{6,7,8}\mu=0}$, i.e., see Eq.(\ref{eqn.01}), extending the deformation space ($\beta_2, \beta_3, \beta_4, \beta_5$) to ($\beta_2, \beta_3, \beta_4, \beta_5, \beta_6, \beta_7, \beta_8$).


With a parameterized nuclear shape, the phenomenological nuclear potential can be calculated. For a nucleus, the Woods-Saxon (WS) potential is more realistic owing to its flat-bottomed and short-range properties. In the project, we numerically solve the Schr\"{o}dinger equation with a deformed WS Hamiltonian~\cite{Dudek1980,Cwiok1987},
\begin{eqnarray}
    \hat{H}_\mathrm{WS}
    &=&
    \hat{T}
    +
    \hat{V}_\mathrm{cent}
    +
    \hat{V}_\mathrm{so} 
    +
   \hat{V}_\mathrm{Coul},
                                                                  \label{eqn.02}
\end{eqnarray}
where the central part of the WS
potential reads
\begin{equation}
    \hat{V}_\mathrm{cent}(\vec{r},{\beta}; V_0,r_0,a_0)
    =
    \frac{V_0[1\pm\kappa(N-Z)/(N+Z)]}{1+\exp[\text{dist}_\Sigma(\vec{r},
    {\beta};r_0)/a_0]},
                                                                  \label{eqn.03}
\end{equation}
where the plus and minus signs hold for protons and neutrons, respectively and the parameter $a_0$ denotes the surface diffuseness. The parameters $V_0$ and $r_0$ represent respectively the central potential depth and central potential radius parameters. The term $\text{dist}_\Sigma(\vec{r},{\beta};r_0)$ represents the distance of a point $\vec{r}$ from the nuclear surface $\Sigma$. The spin-orbit potential, which strongly affects the level order and depends on the gradient of the central potential with new parameters, is defined by
\begin{eqnarray}
     \hat{V}_\mathrm{so}(\vec{r},\hat{p},\hat{s},{\beta};\lambda, r_\mathrm{so}, a_\mathrm{so})
  =  
    -\lambda
    \Big[
    \frac{\hbar}{2mc}
    \Big]^2      
     \nabla V^\mathrm{so}
    \times\hat{p}\cdot\hat{s},  
                                                                  \label{eqn.04}
\end{eqnarray}
where 
\begin{eqnarray}
V^\mathrm{so}(\vec{r},{\beta}; r_\mathrm{so}, a_\mathrm{so}) = \frac{V_0[1\pm\kappa(N-Z)/(N+Z)]}{1
    +
    \mathrm{exp}[\mathrm{dist}_{\Sigma_\mathrm{so}}(\vec{r},{\beta};r_\mathrm{so})/a_\mathrm{so}]},
                                                                  \label{eqn.05}
\end{eqnarray}
and the parameter $\lambda$ denotes the strength of the effective spin-orbit force acting on the individual nucleons. It should be stressed that the new surface $\Sigma_{so}$ is different from the one in Eq.~(\ref{eqn.05}) due to the different radius parameter $r_{so}$. Also, the spin-orbit diffusivity parameter $a_{so}$ is usually updated. 
For protons, a classical electrostatic potential of a uniformly charged drop is used for describing the Coulomb potential, which is defined as
\begin{eqnarray}
V^\mathrm{Coul}(\vec{r},{\beta}) = Ze\int_{\Sigma} \frac{d^3\vec{r^\prime}}{|\vec{r}-\vec{r^\prime}|} ,
                                                                  \label{eqn.06}
\end{eqnarray}
where the integration extends over the volume delimited by the surface $\Sigma$.


During the process of calculating the WS Hamiltonian matrix, we use the eigenfunctions of the axially deformed harmonic oscillator potential in the cylindrical coordinate system as the basis function, as seen below,
\begin{equation}
|n_\rho n_z \Lambda \Sigma \rangle
  =
  \psi_{n_\rho}^\Lambda(\rho)\psi_{n_z}(z)
  \psi_{\Lambda}(\varphi)\chi(\Sigma).
                                                                  \label{eqn.07}
\end{equation}
For more details, one can see e.g., Ref.~\cite{Cwiok1987}. Note that the eigenfunctions with $N\leqslant 12$ and 14 are chosen as the basis set for protons and neutrons, respectively. The corresponding single-particle levels (eigenvalues) and wave functions (eigenvectors) will be obtained by diagonalizing the Hamiltonian matrix. It is found that, by such a cutoff, the calculated results (e.g., single-particle energies) are sufficiently stable with respect to a possible enlargement of the basis space.

Based on the obtained single-particle levels at the corresponding nuclear shape, the quantum shell-correction and pairing-energy contributions can be further calculated by the Strutinsky method~\cite{Bolsterli1972} and the Lipkin-Nogami (LN) method~\cite{Pradhan1973,Satula1994NPA}. In the these methods, the microscopic shell-correction energy is given by
\begin{equation}
\delta E_\mathrm{shell}(Z,N,\hat{\beta})=\sum e_i -\int e\tilde{g}(e)de,
                                                    \label{eqn.08}
\end{equation}
where $e_i$ denotes the calculated single-particle levels and $\tilde{g}(e)$ is the so-called smooth level density. The smoothed distribution function $\tilde{g}(e)$ was early defined as,
\begin{equation}
\tilde{g}(e,\gamma)\equiv \frac{1}{\gamma\sqrt{\pi}}
                   \sum_i {\rm exp}[-\frac{(e-e_i)^2}{\gamma^2}],
                                                    \label{eqn.09}
\end{equation}
where $\gamma$ denotes the smoothing parameter. To eliminate any possibly strong dependence of the $\gamma$-parameter, the level density $\tilde{g}(e)$, optimized by a curvature-correction polynomial
$P_p(x)$, is usually given by~\cite{Werner1995,Nilsson1969,Strutinsky1975,
Bolsterli1972},
\begin{equation}
\tilde{g}(e,\gamma, p) = \frac{1}{\gamma\sqrt{\pi}}
                   \sum_{i=1} P_p(\frac{e-e_i}{\gamma})
                   \times{\rm exp}[-\frac{(e-e_i)^2}{\gamma^2}].
                                                    \label{eqn.10}
\end{equation}
The corrective polynomial $P_p(x)$ can be expanded in terms of the Hermite or Laguerre polynomials. The expanded coefficients can be obtained by using the orthogonality properties of these polynomials and Strutinsky condition~\cite{Pomorski2004}. In the present work, a sixth-order Hermite polynomial and a smoothing parameter $\gamma
=1.20\hbar \omega _0$, where $\hbar \omega_0=41/A^{1/3}$ MeV, are adopted~\cite{Bolsterli1972}.

Besides the shell correction, another important quantum correction is the pairing-energy contribution. It deserves noting that there exist various variants of the pairing energy in the MM approach~\cite{Gaamouci2021}. Several kinds of the phenomenological expressions are widely adopted, such as pairing correlation and pairing correction energies employing or not employing the particle number projection technique. We employ the
LN method, an approximately particle number projection technique, to treat the pairing-energy calculation~\cite{Pradhan1973,Satula1994NPA}. Such a pairing treatment can help avoiding
not only the spurious pairing phase transition but also particle number fluctuation encountered in the simpler BCS calculation. In this method, the LN pairing energy for an even-even nucleon system at ``paired solution'' (pairing gap $\Delta\neq 0$) is caclulated by~\cite{Pradhan1973,Moller1995}
\begin{eqnarray}
E_\mathrm{LN}&=&\sum_{k}2{v_k}^2e_k-\frac{{\Delta}^2}G-G\sum_{k}{v_k}^4
\nonumber \\
     && -4{\lambda}_2\sum_{k}{u_k}^2{v_k}^2,
                                                                  \label{eqn.11}
\end{eqnarray}
where ${v_k}^2$, $e_k$, $\Delta$ and ${\lambda}_2$ represent the occupation probabilities, single-particle energies, pairing gap and number-fluctuation constant, respectively. The monopole pairing strength $G$ is determined by the average gap method~\cite{Xu98}. For the case of ``no-pairing solution'' ($\Delta = 0$), its partner expression is
\begin{eqnarray}
E_\mathrm{LN}(\Delta=0)&=&\sum_{k}2e_k-G\frac{N}{2}.
                                                                  \label{eqn.12}
\end{eqnarray}
The difference between paired solution $E_{LN}$ and no-pairing solution $E_{LN}$($\Delta=0$) is usually referred to as the pairing correlation, which can be written as,
\begin{eqnarray}
\delta E_\mathrm{pair}&=&\sum_{k}2{v_k}^2e_k-\frac{{\Delta}^2}G-G\sum_{k}{v_k}^4
\nonumber \\
     && -4{\lambda}_2\sum_{k}{u_k}^2{v_k}^2+G\frac{N}{2}-\sum_{k}2e_k. \nonumber \\
                                                                  \label{eqn.13}
\end{eqnarray}

Following Refs.~\cite{Moller1995,Moller2016}, we define the total microscopic energy as, 
\begin{eqnarray}
    E_\mathrm{micro}(Z,N,{\beta})
    &=&
    \delta E_\mathrm{shell}(Z,N,{\beta}) 
    +
    \delta E_\mathrm{pair}(Z,N,{\beta}). \nonumber \\
                                                                  \label{eqn.14}
\end{eqnarray}
Such a definition is equivalent to the concept of ``shell correction'' $\delta E_{shell}$ ($\equiv E_{LN} - \tilde{E}_{Strut}$), cf. Eq. (1) in Ref.~\cite{Xu98}, merging the quantum shell correction and pairing contribution. For clarifying some confusing points, it should be pointed out that, in Ref.~\cite{Xu98}, the definition of $E_{LN}$ includes the term $G\frac{N}{2}$. {It should be mentioned that the microscopic energy includes the proton and neutron contributions simultaneously.} 

Using the standard liquid-drop model~\cite{Myers66}, the macroscopic energy can be calculated. Since we pay attention to the deformation effects instead of, e.g., masses, in the PES calculation, the deformation liquid-drop energy {(relative to the spherical liquid drop)} is adopted ~\cite{Cwiok1987,Myers66,Bolsterli1972}, as seen below,
\begin{eqnarray}
E_{\mathrm{macro}}(Z,N,\beta)
&=& 
    E_{\mathrm{ld}}(Z,N,{\beta}) \nonumber \\
&=&
\{[B_{\mathrm{S}}(\beta)-1]
+
2\chi[B_{\mathrm{C}}(\beta)-1]\}E_{\mathrm{S}}^{(0)},  \nonumber \\
                                                                  \label{eqn.15}
\end{eqnarray}
where the spherical surface energy $E_{S}^{(0)}$ and the fissility parameter $\chi$ are Z and N dependent, cf. Refs.~\cite{Myers66,Bolsterli1972}. The relative surface and Coulomb energies $B_{S}$ and $B_{C}$ are only functions of nuclear shape.

Within the framework of MM model, the total energy can be calculated by~\cite{Dudek1988,Moller1995},
\begin{eqnarray}
    E_{\mathrm{total}}(Z,N,{\beta})
    &=&
    E_{\mathrm{macro}}(Z,N,{\beta})                   
    +
    E_{\mathrm{micro}}(Z,N,{\beta}). \nonumber \\  
                                                                  \label{eqn.16}
\end{eqnarray}
Once the total energy is obtained at each sampling deformation grid, we can obtain the smoothly potential-energy surface/map with the help of the interpolation techniques, e.g., a spline function, and then investigate nuclear properties including the equilibrium deformations, shape co-existance, fission path and other physical quantities/processes.

\section{Results and Discussion}
\label{Results and Discussion}
In this project, we restrict ourselves to axially symmetric shapes and perform the numerical calculations on a seven-dimensional deformation space. Namely, we calculate (25, 13, 13, 7, 5, 5, 5) points for ($\beta_2,\beta_3, \beta_4,\beta_5, \beta_6,\beta_7, \beta_8)$ deformations, respectively, taking the size 0.05 as the deformation step. Our primary concern is how the high-order multipolarity deformations and their couplings affect the potential energy landscapes.

\begin{figure}[htbp]
	\centering
	\includegraphics[width=0.75\linewidth]{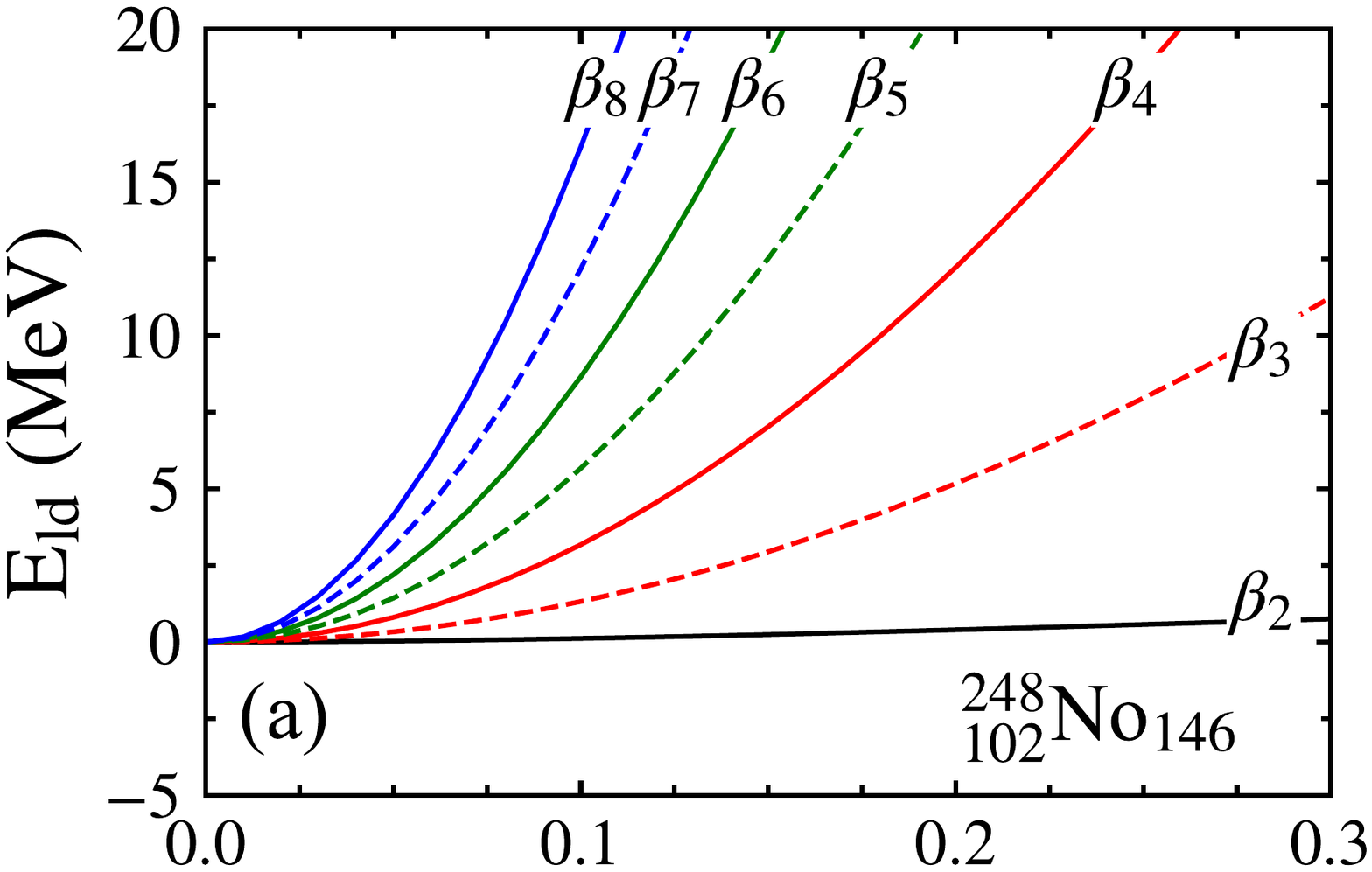}\\
	\vspace{-0.5cm}
	\includegraphics[width=0.75\linewidth]{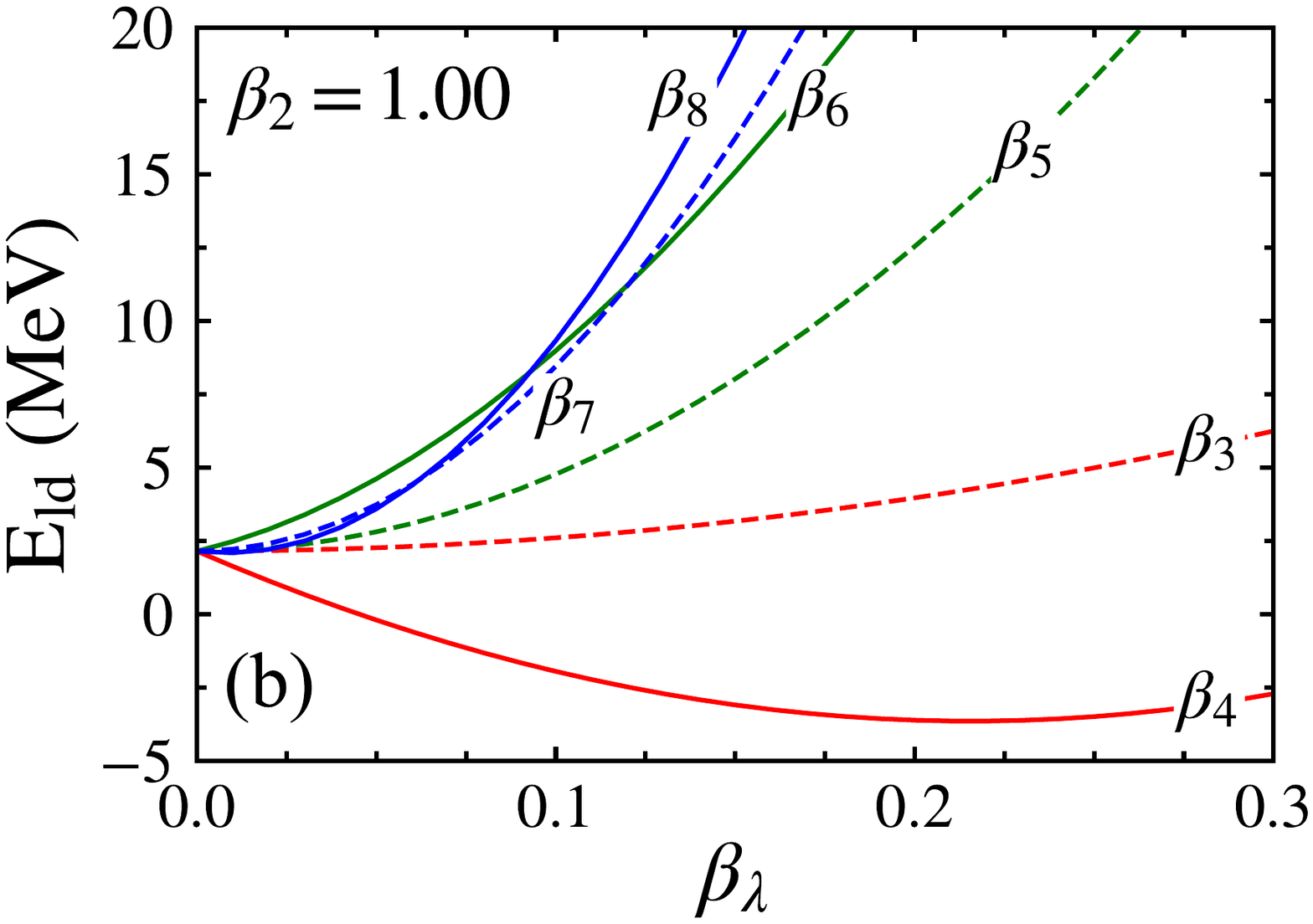}
	\vspace{-0.2cm}
	\caption{Macroscopic deformation energies as functions of separate deformations ${\beta_\lambda}$ for $^{248}$No, $\lambda = 2, 3, 4, 5, 6, 7, 8$. Note that the $\beta_2$ deformation is fixed to 1.0 in subfigue (b).}
	\label{Fig.1}
\end{figure}

For the heavy nuclear systems, we know that nuclear stability is approximately governed by the competition between the surface tension of the nuclear liquid drop and the strong Coulomb repulsion between the numerous protons. The former tends to hold the system together, while the latter drives the nucleus towards spontaneous fission. To understand the influence of different deformation parameters on macroscopic energy, taking the $^{248}$No as an example, the evolution of deformed liquid-drop energies near the spherical and elongated shapes is illustrated in Fig.~\ref{Fig.1}. Such a heavy nucleus, the macroscopic liqui-drop energy almost keep a small constant with the changing $\beta_2$, agreeing with our previous study~\cite{Li2023}. One can notice that the nuclear stiffness (usually, defined by $\partial E_{ld}/\partial \beta_\lambda$) near the spherical shape increases with the increasing $\lambda$, as seen in Fig.~\ref{Fig.1}(a), indicating that it is more difficult to occur high-order deformation. However, at the elongated case, e.g., $\beta_2=1.0$, cf. Fig.~\ref{Fig.1}(b), the nuclear shape relatively becomes soft along the direction of each deformation degree of freedom. In particular, it can be seen from Fig.~\ref{Fig.1}(b) that the stable $\beta_4$ deformation appears under the circumstances. Even, the stiffnesses along $\beta_7$ and $\beta_8$ are almost same and smaller than that along the lower-order deformation $\beta_6$ at about $\beta_{6,7,8} < 0.1$, which means that high-order $\beta_{7,8}$ deformations may be more favored than $\beta_6$. In the practical calculations, the different couplings of different deformations may be complex. 

 
%

\begin{figure}[htbp]
	\centering
	\includegraphics[width=0.45\linewidth]{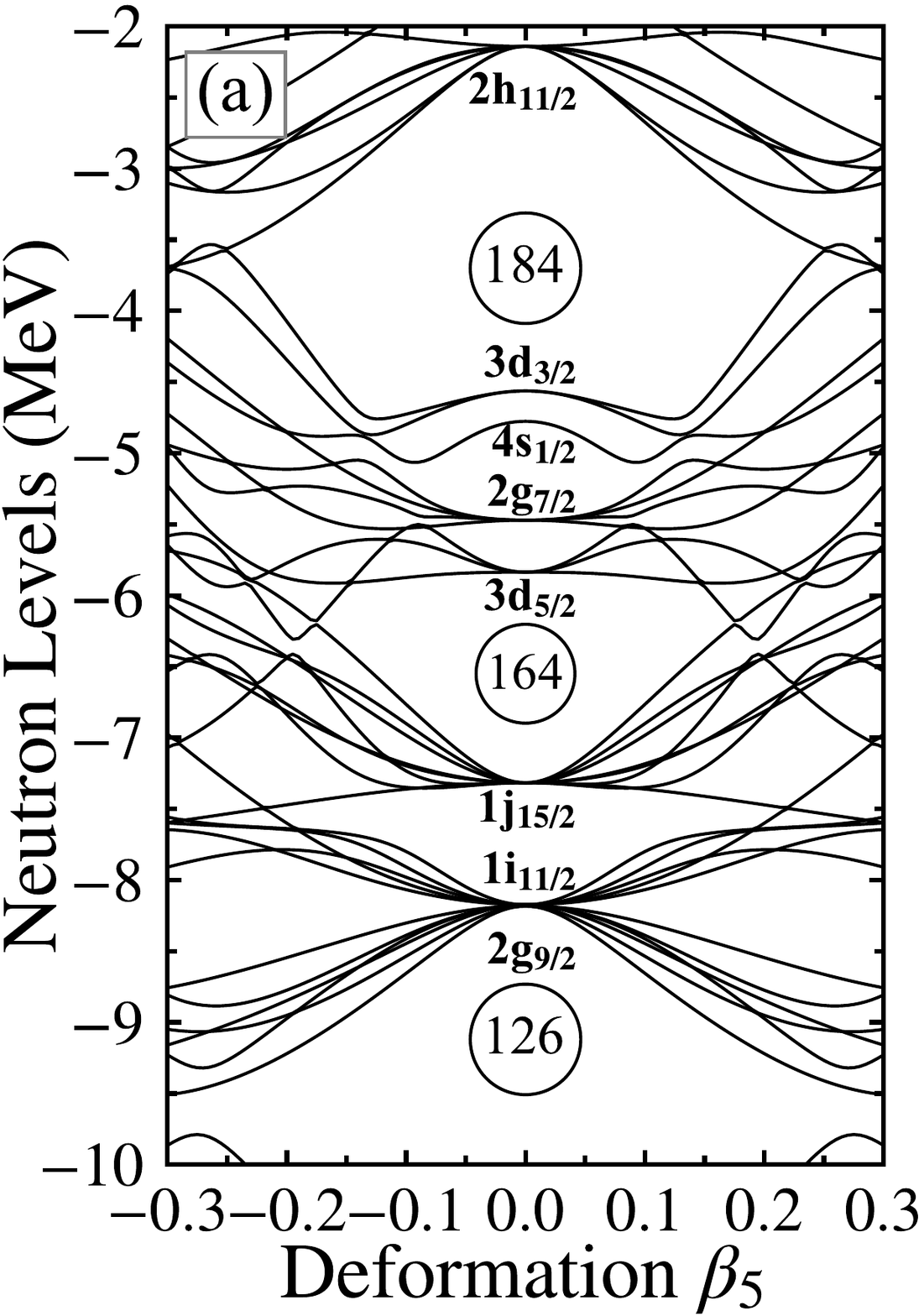}
	\includegraphics[width=0.45\linewidth]{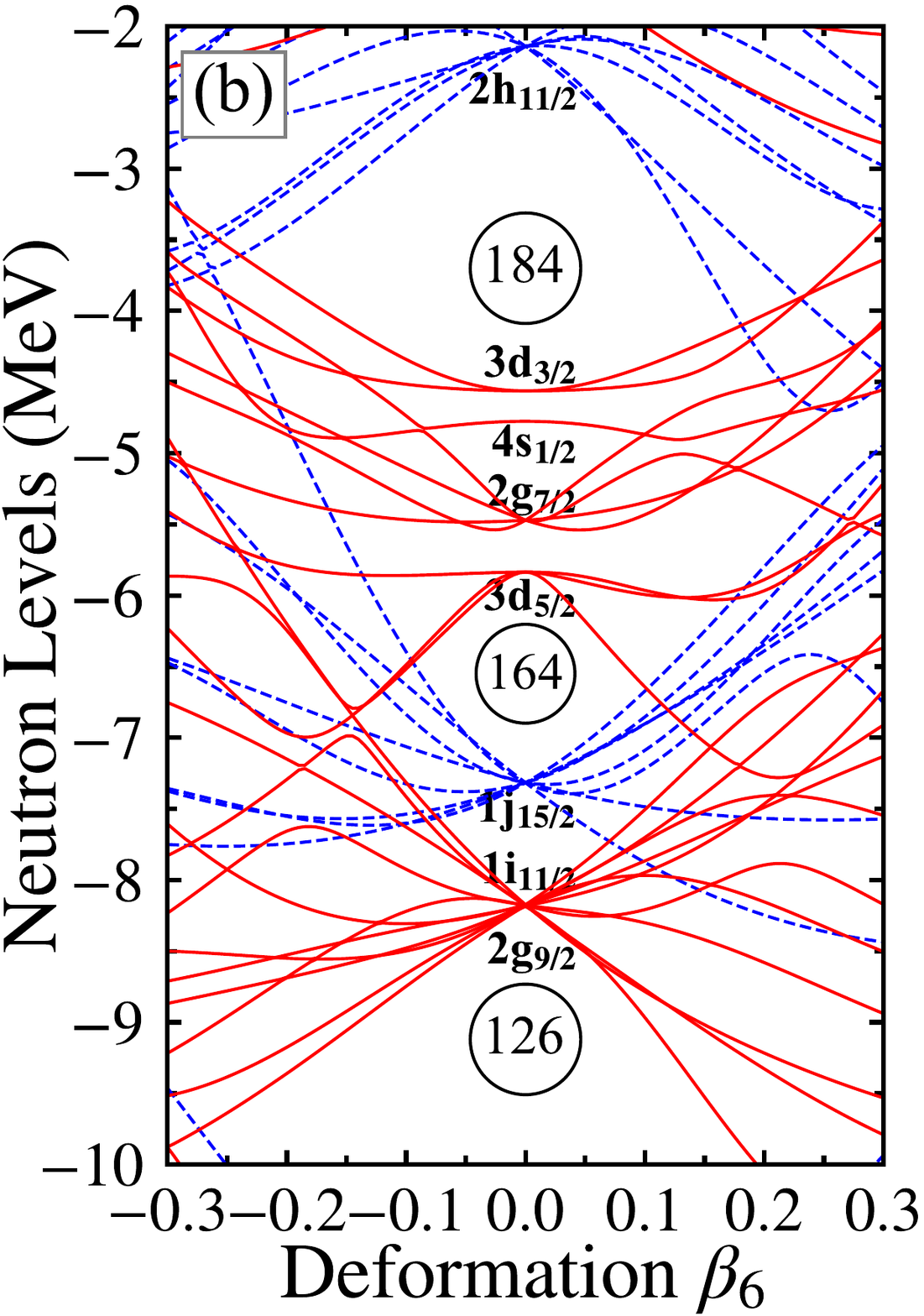}
	\includegraphics[width=0.45\linewidth]{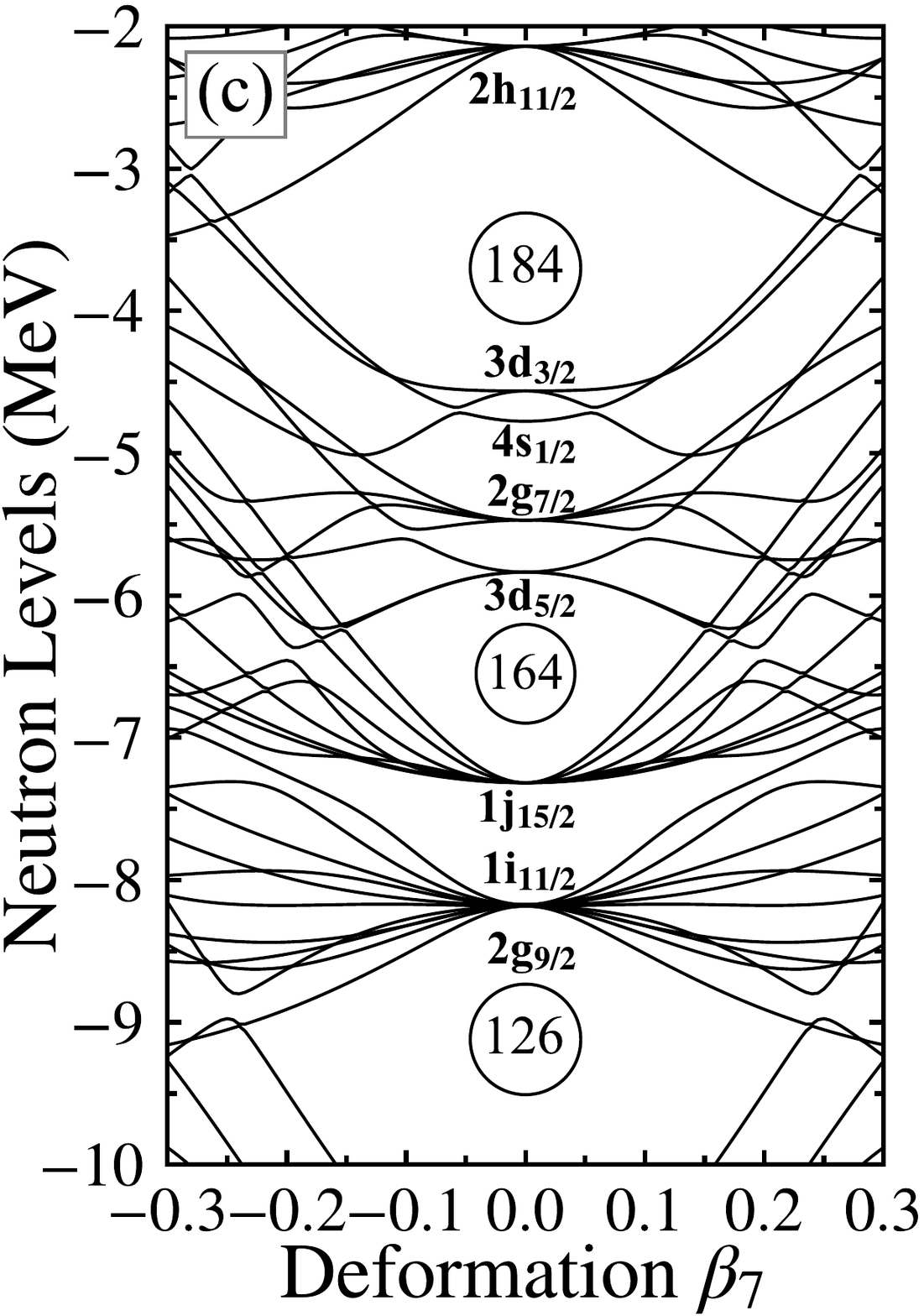}
	\includegraphics[width=0.45\linewidth]{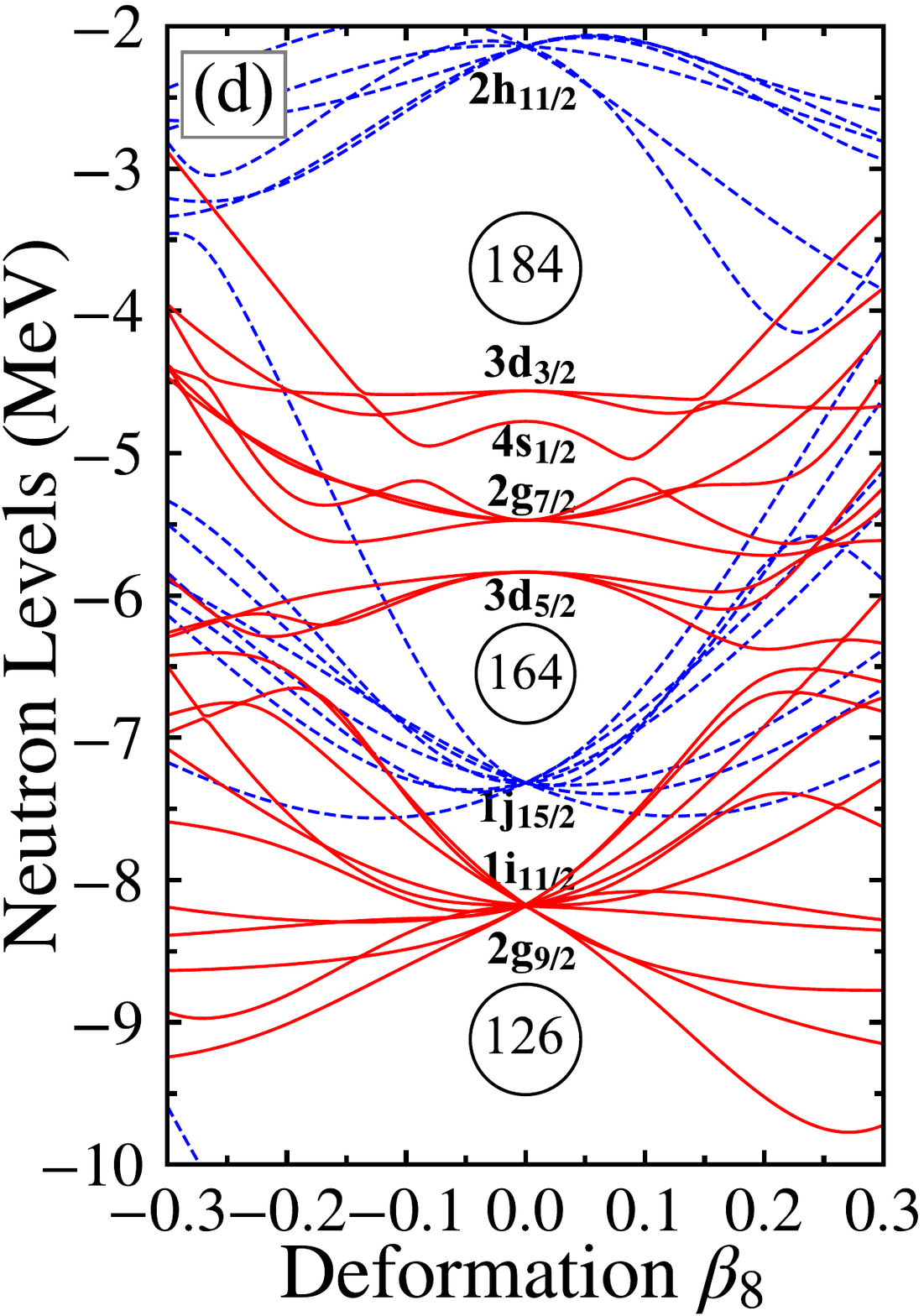}
	\caption{Neutron single-particle energies as functions of separate deformation $\beta_5$ (a), $\beta_6$ (b), $\beta_7$ (c) and $\beta_8$ (d) for $^{248}$No. { For each subplot, other deformation parameters are set to zero and the spherical quantum numbers $nlj$ are given as labels.} In (b) and (d), the solid red and dash blue lines indicate the positive and negative-parity levels, respectively.}
	\label{Fig.2}
\end{figure}

\begin{figure*}[htbp]
	\centering
	\includegraphics[width=0.4\linewidth]{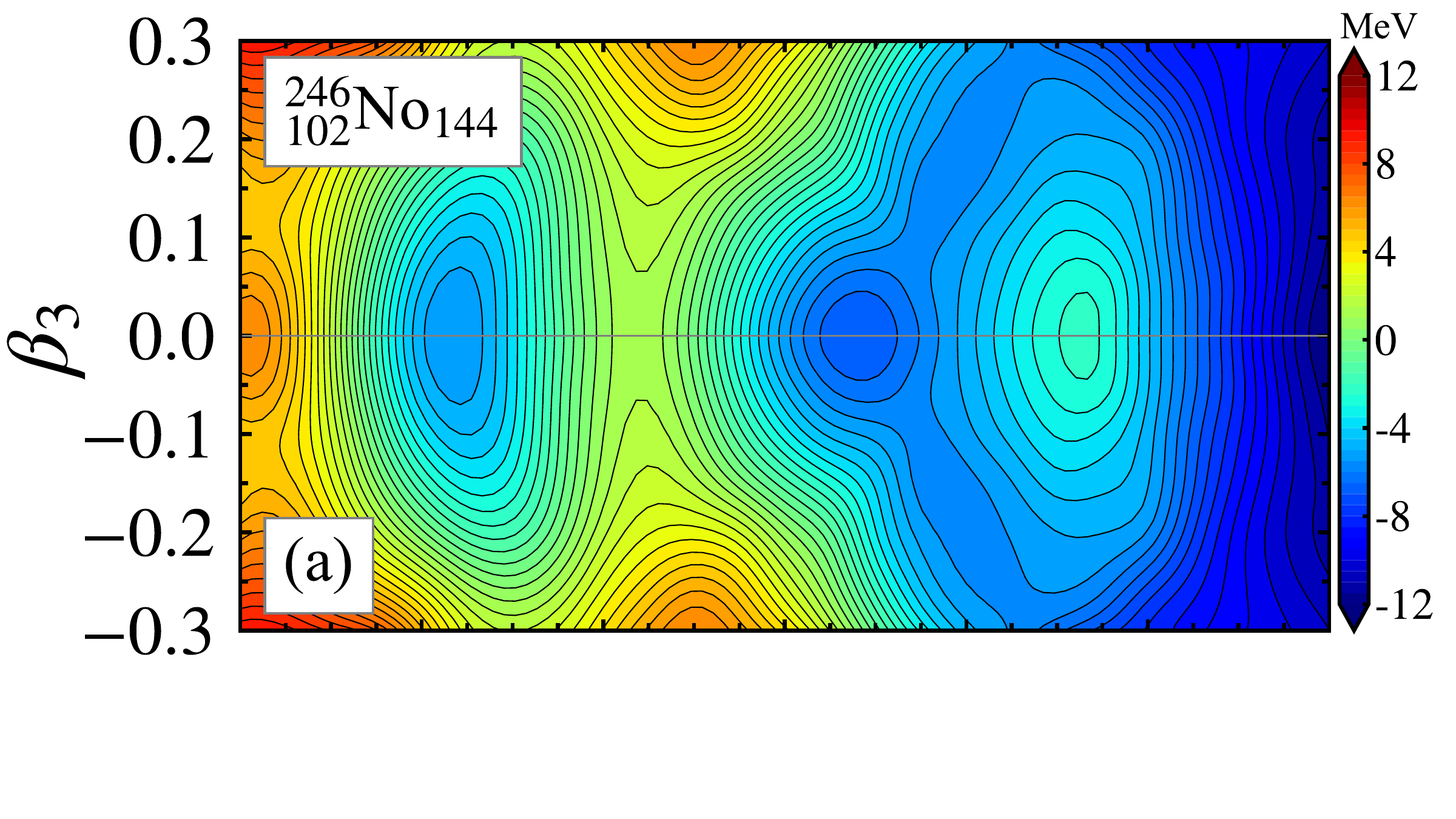}
	\includegraphics[width=0.4\linewidth]{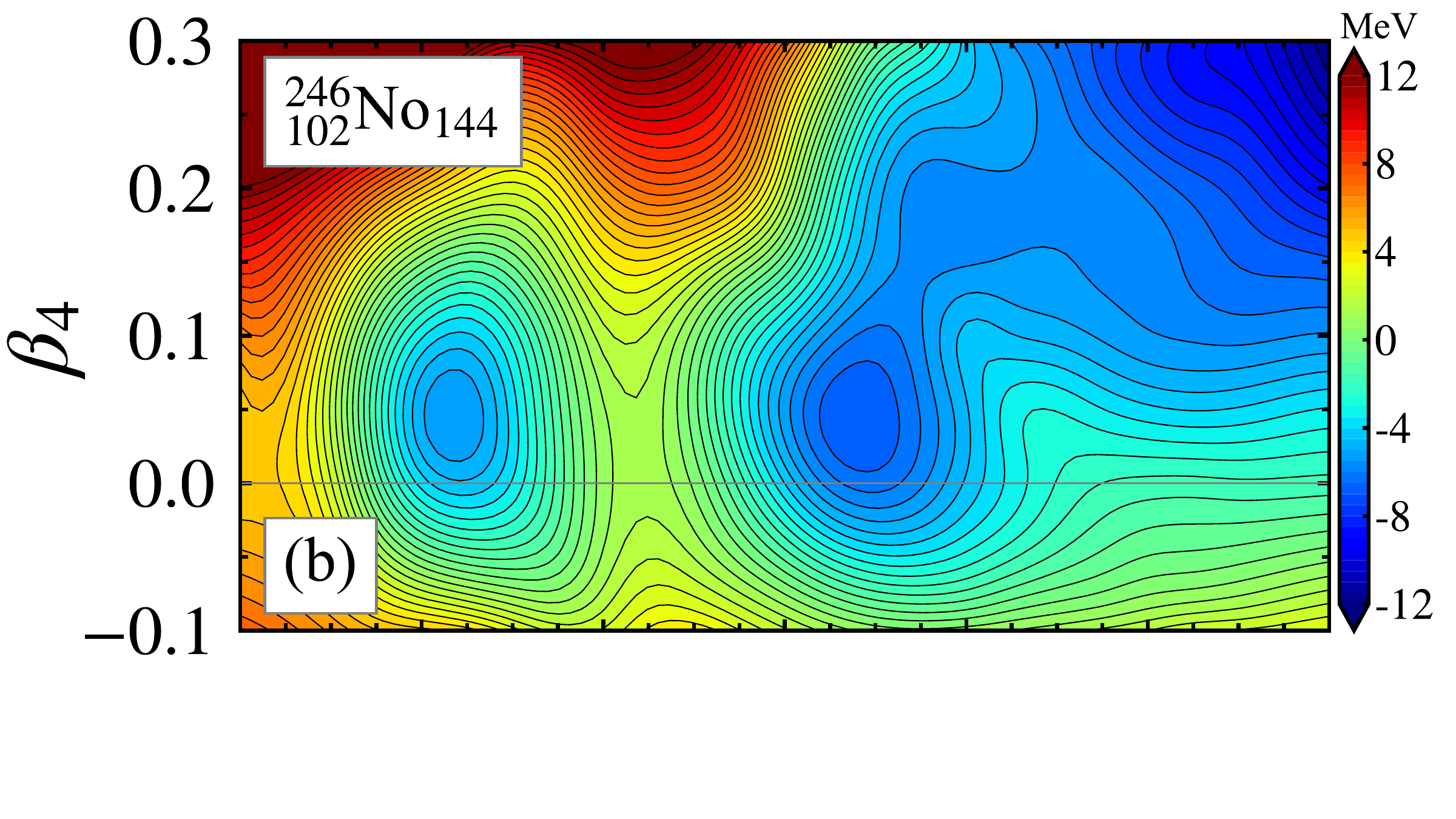}\\
	\vspace{-0.8cm}
	\includegraphics[width=0.4\linewidth]{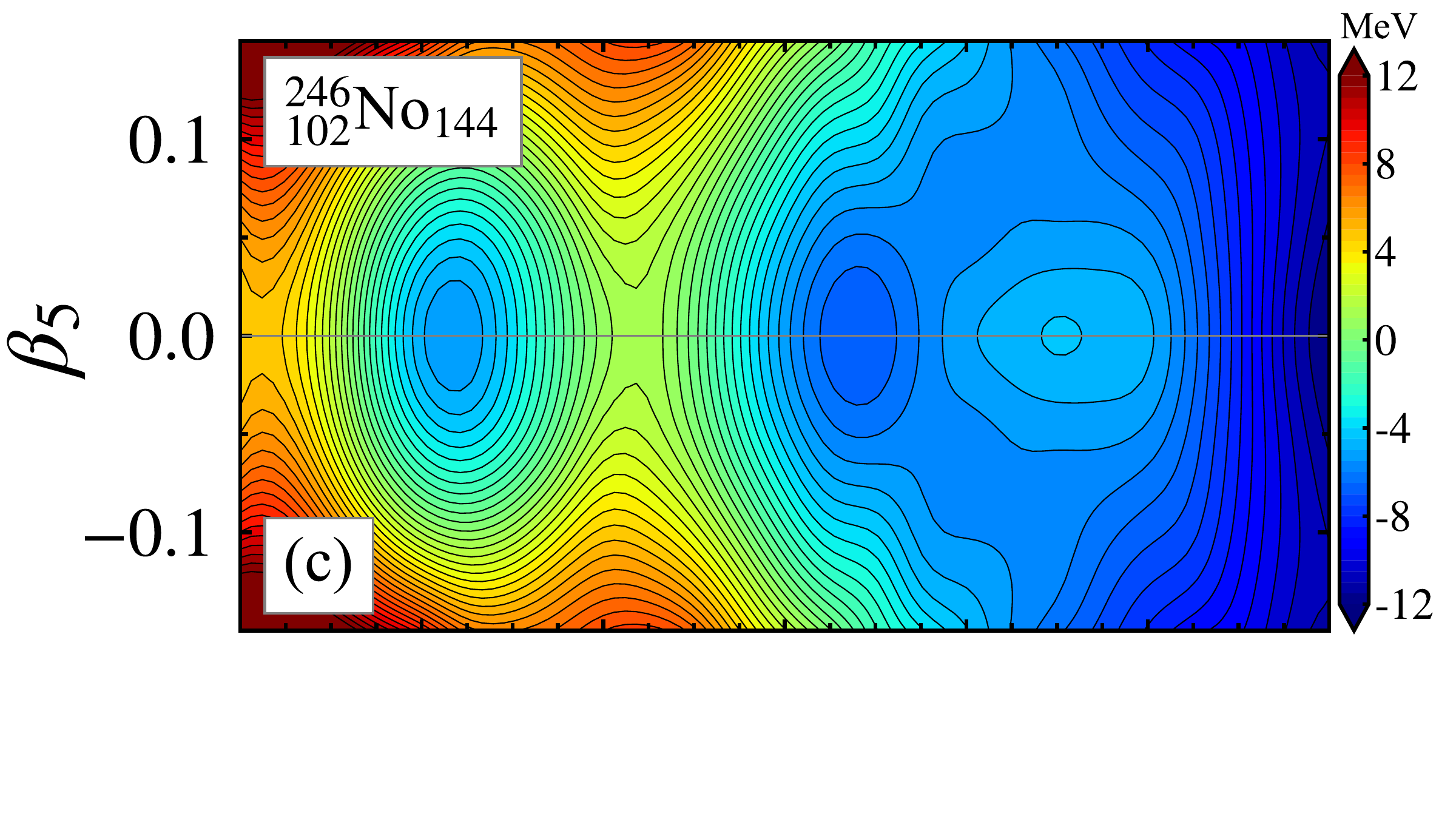}
	\includegraphics[width=0.4\linewidth]{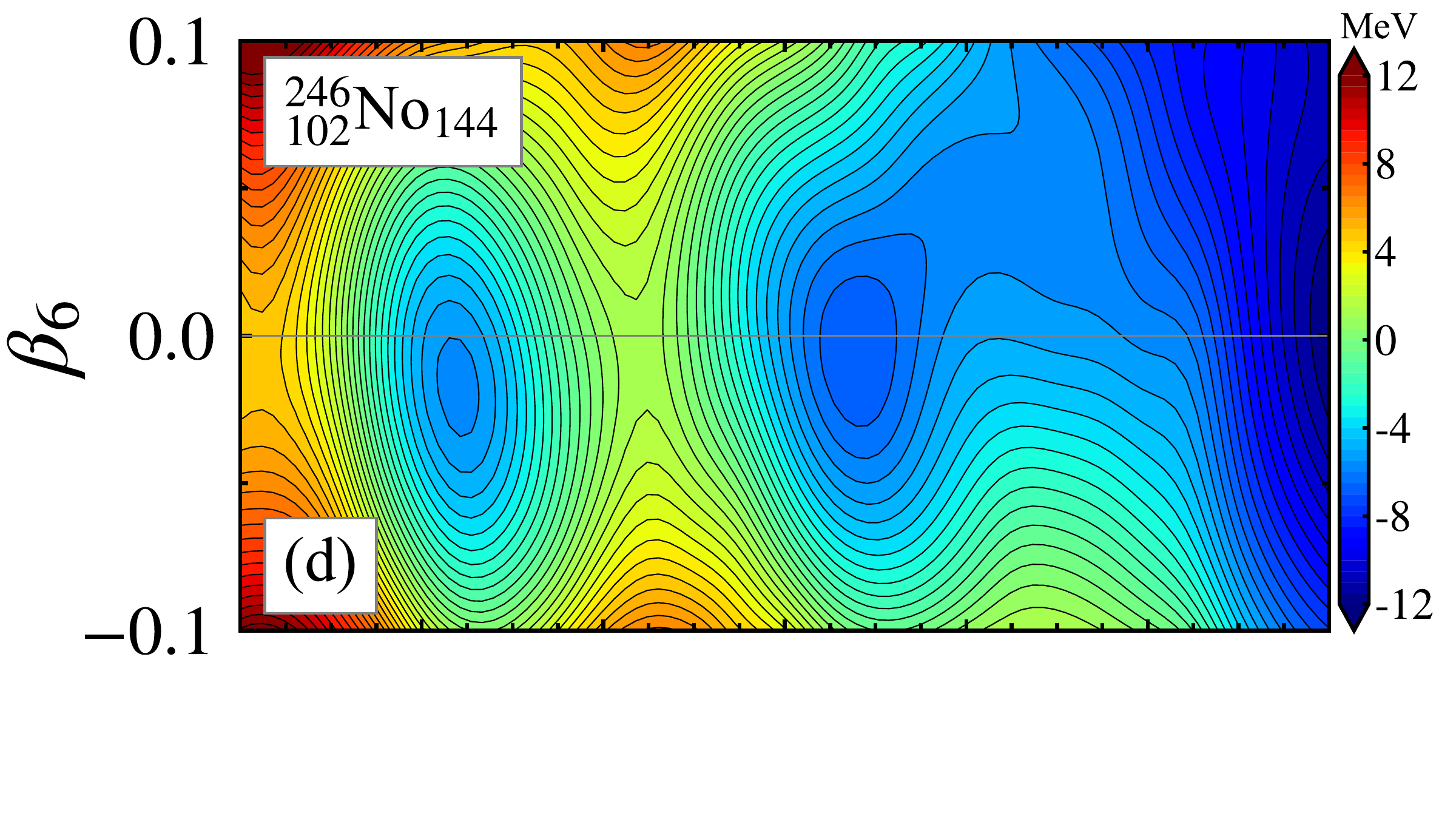}\\
	\vspace{-0.8cm}
	\includegraphics[width=0.4\linewidth]{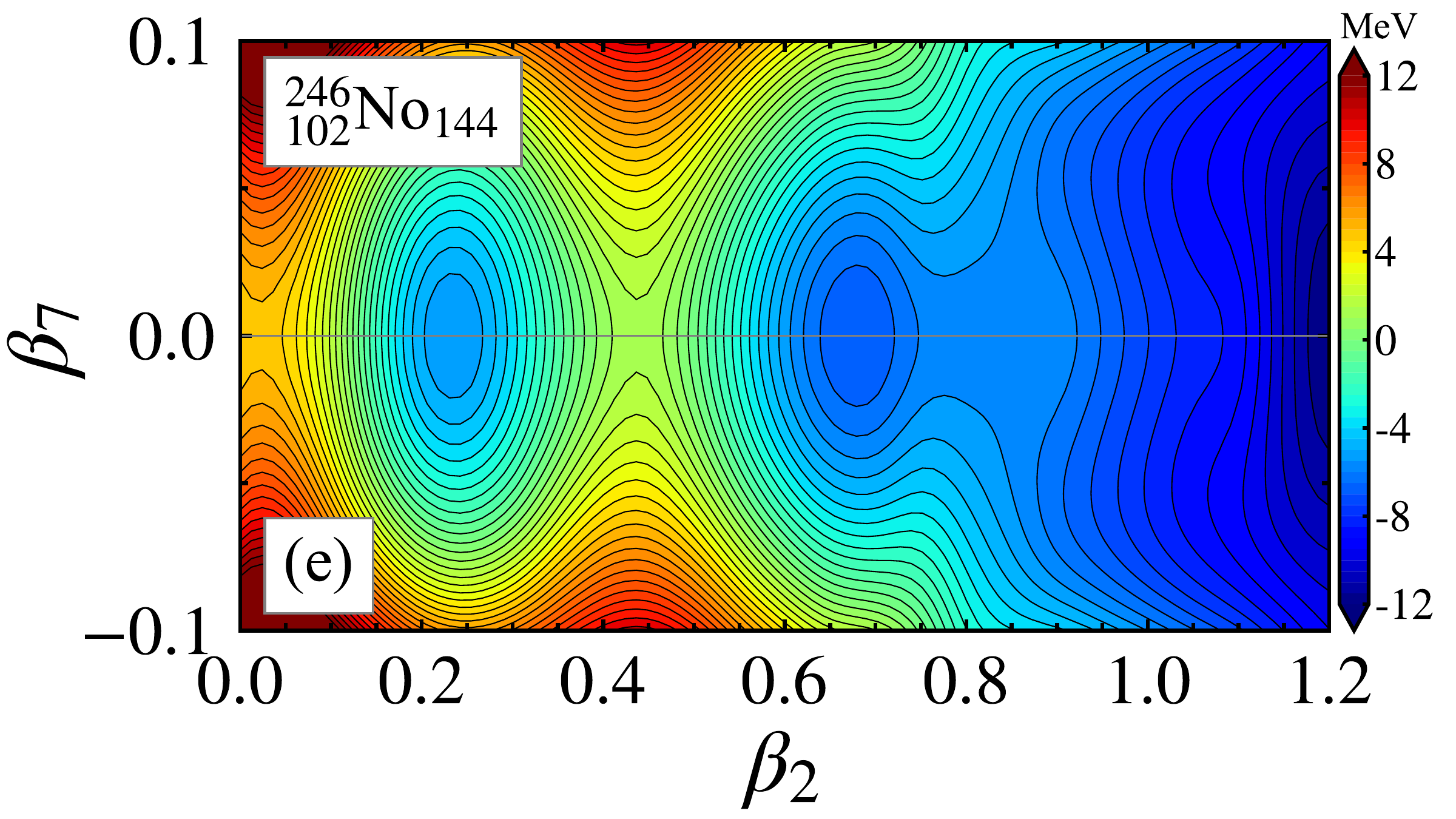}
	\includegraphics[width=0.4\linewidth]{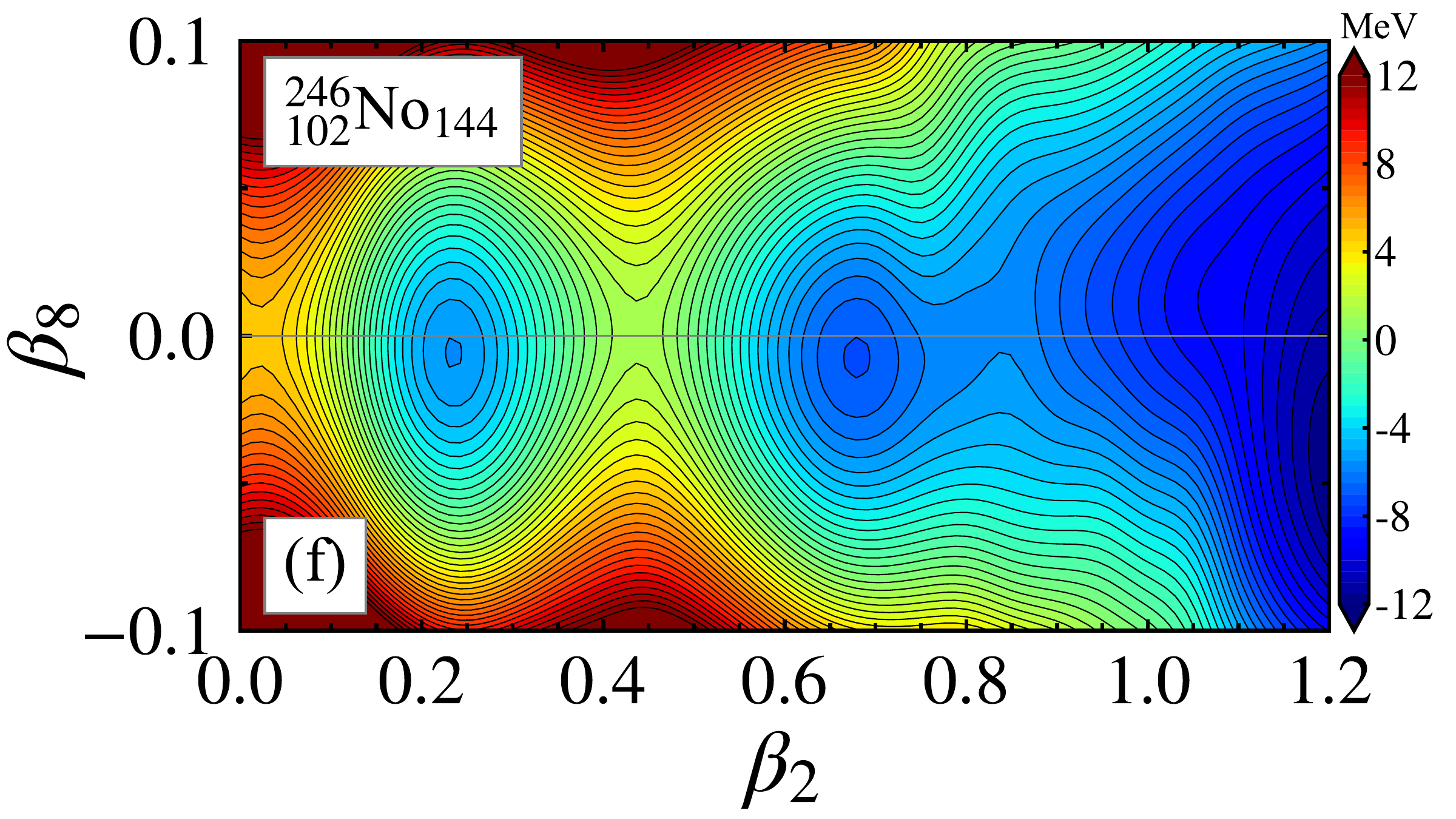}
	\caption{Potential-energy projections on ($\beta_2$, $\beta_{\lambda =3,4,5,6,7,8}$) planes, contour-line separation of 0.5 MeV, minimized at each deformation point over other deformations, for the $^{246}$No nucleus. For more details see the text.}
	\label{Fig.3}
	\vspace{0.5cm}
	\centering
	\includegraphics[width=0.4\linewidth]{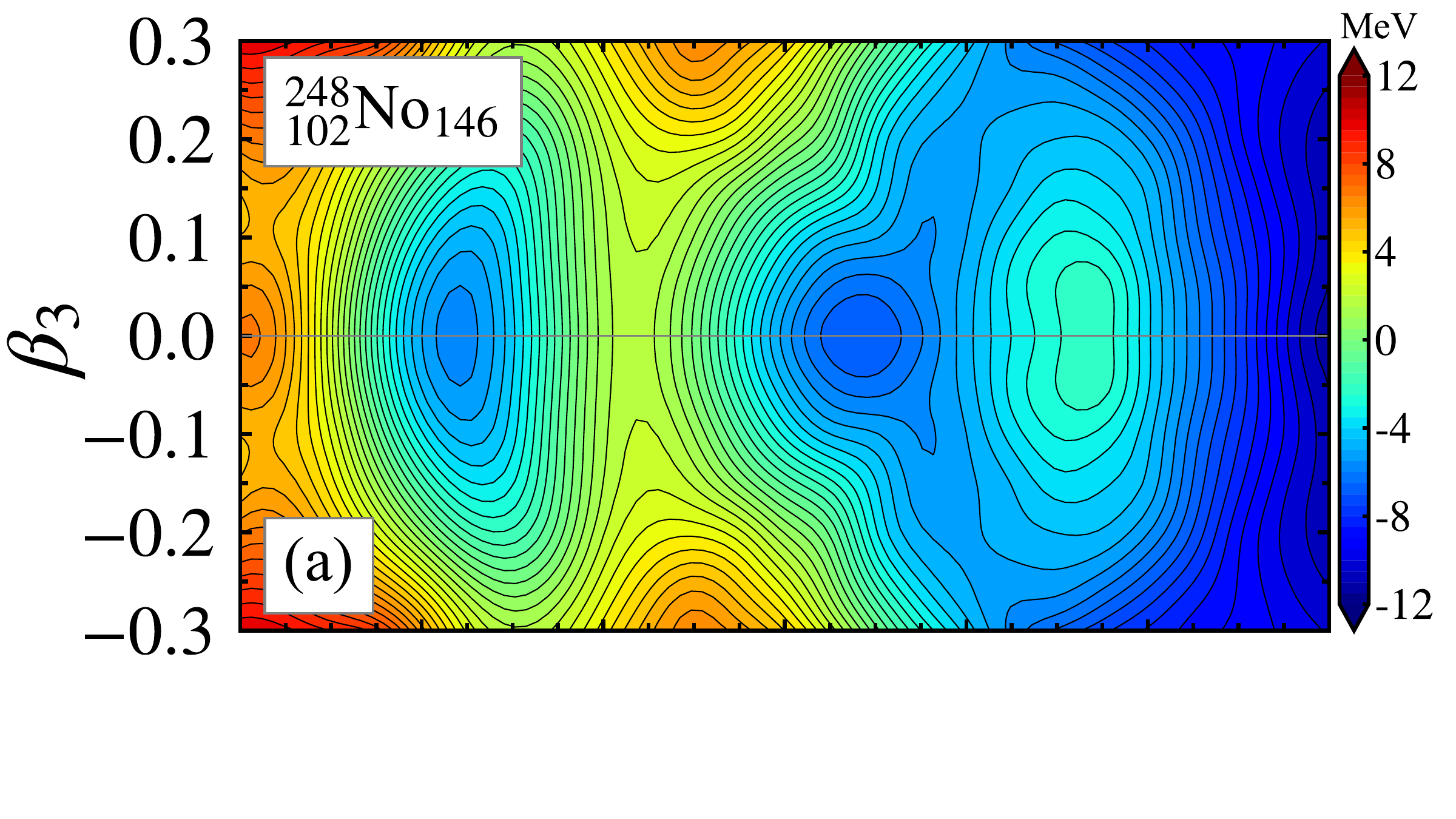}
	\includegraphics[width=0.4\linewidth]{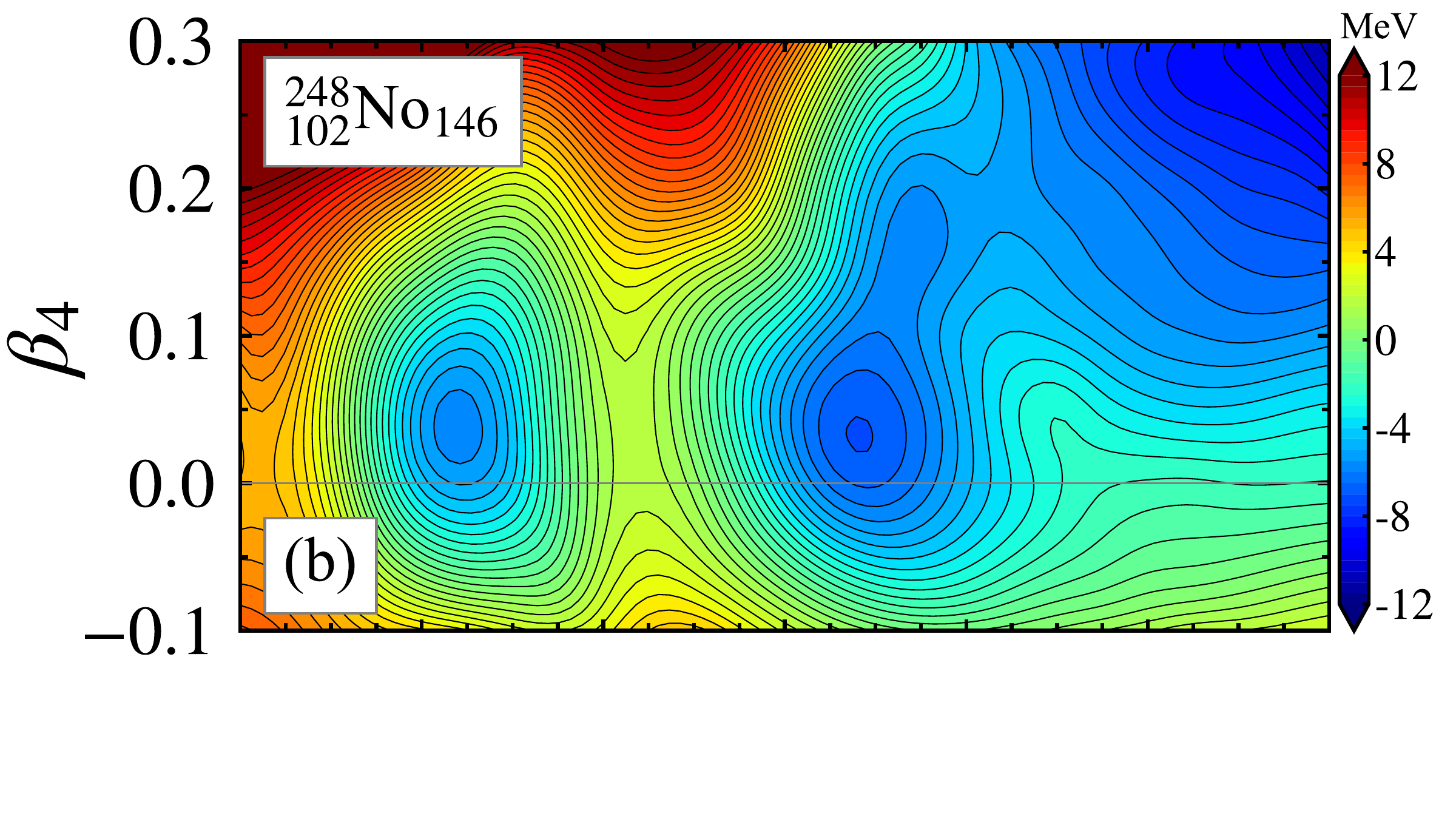}\\
	\vspace{-0.8cm}	
	\includegraphics[width=0.4\linewidth]{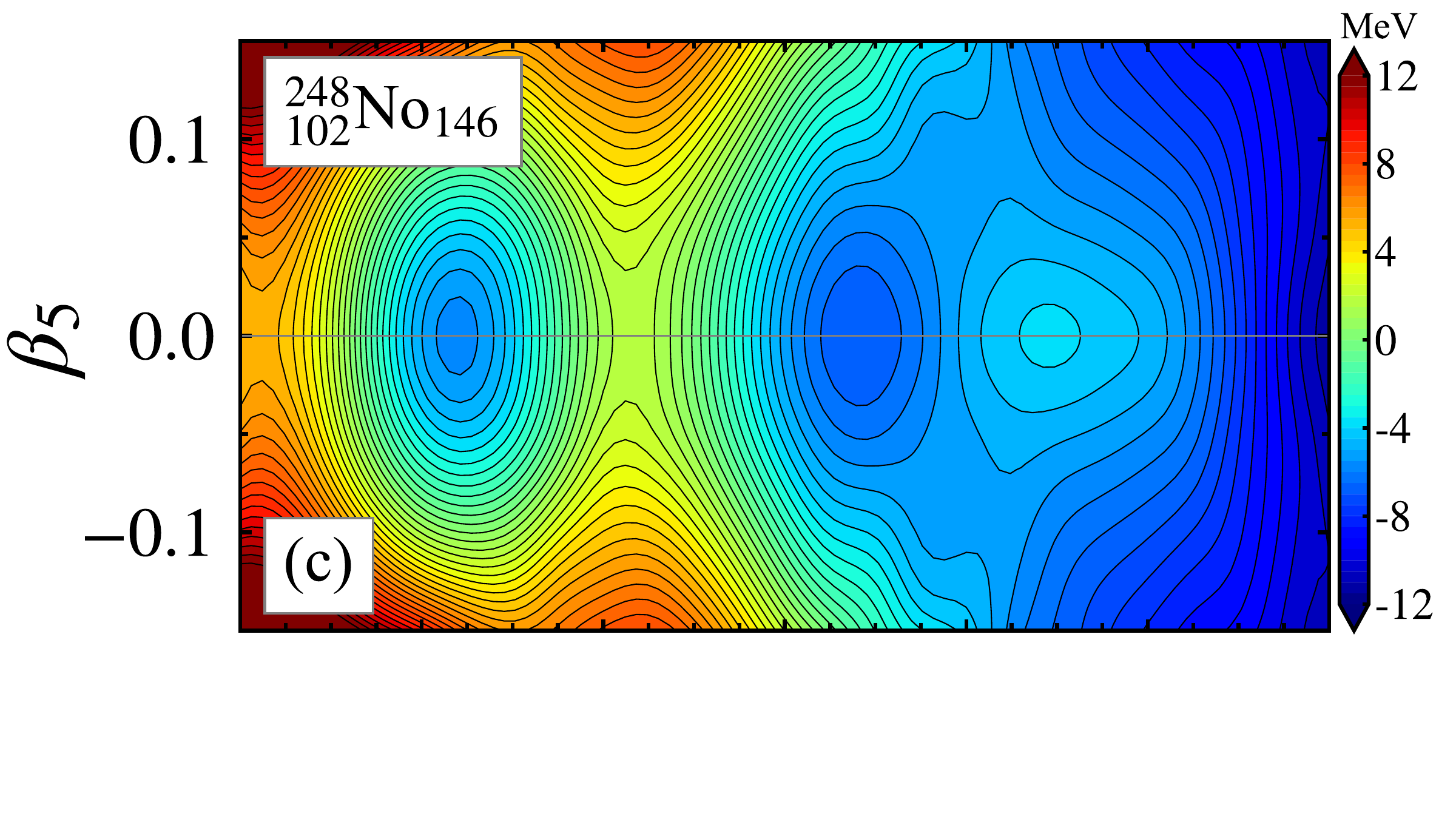}
	\includegraphics[width=0.4\linewidth]{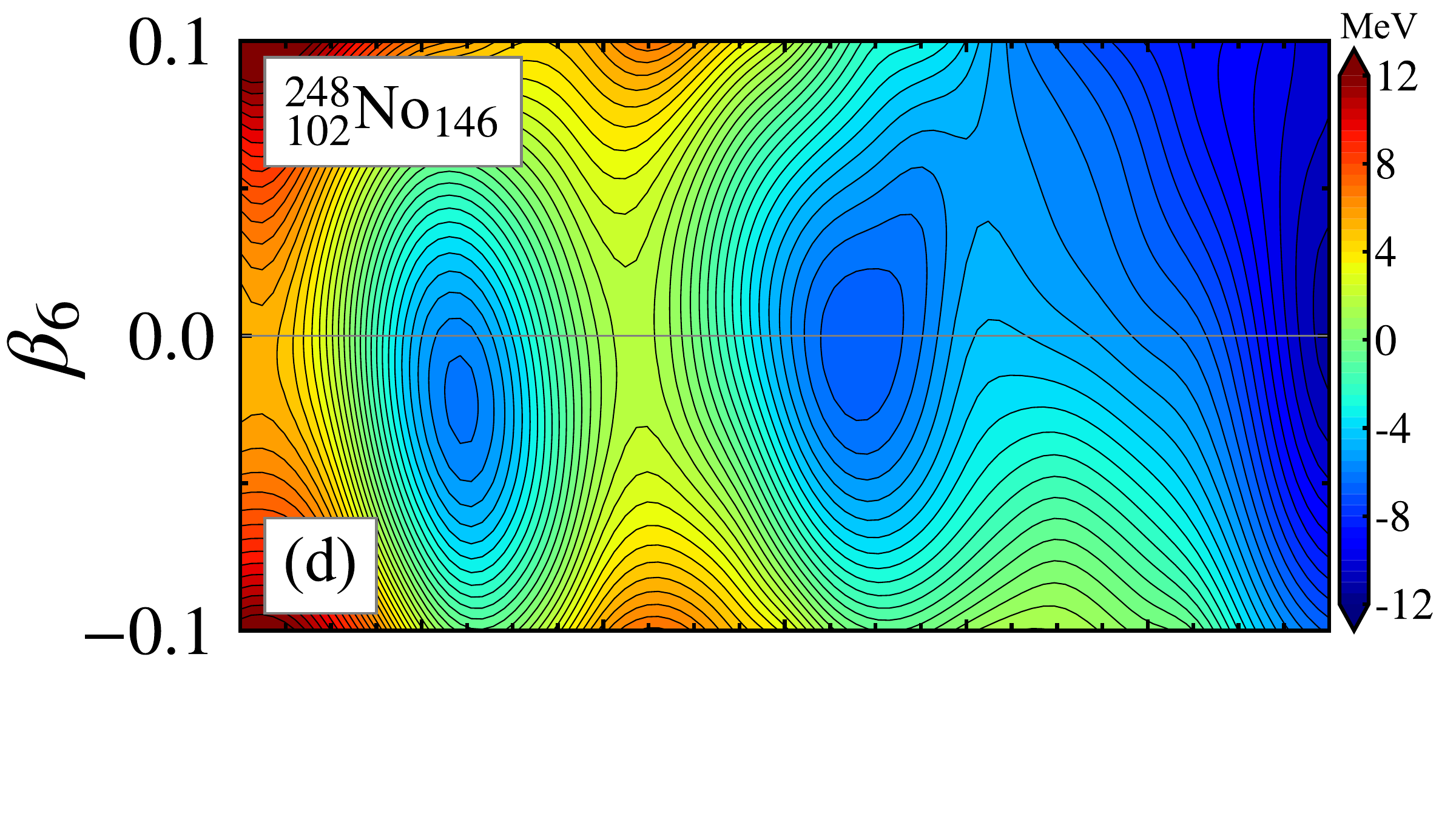}\\
	\vspace{-0.8cm}
	\includegraphics[width=0.4\linewidth]{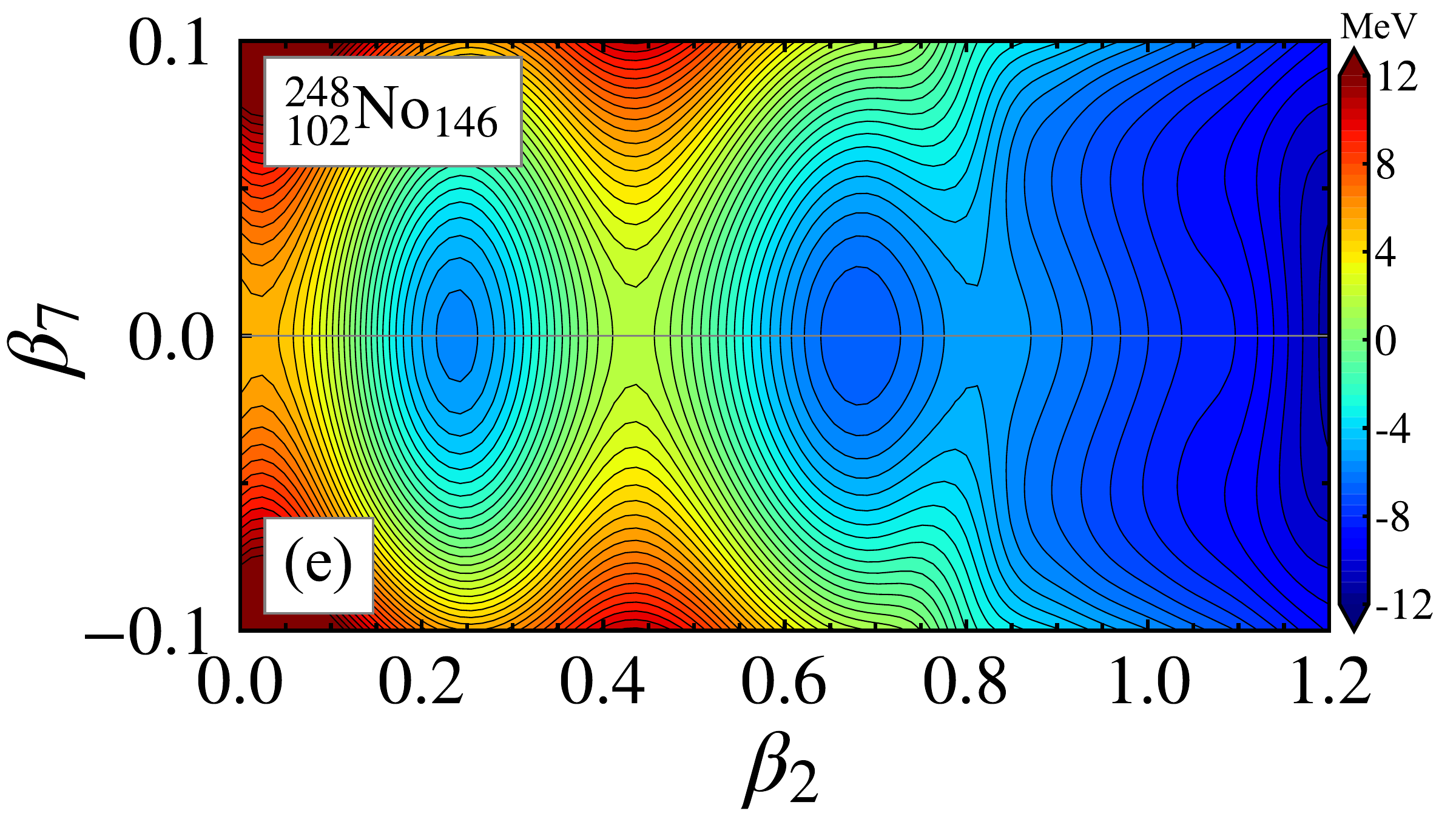}
	\includegraphics[width=0.4\linewidth]{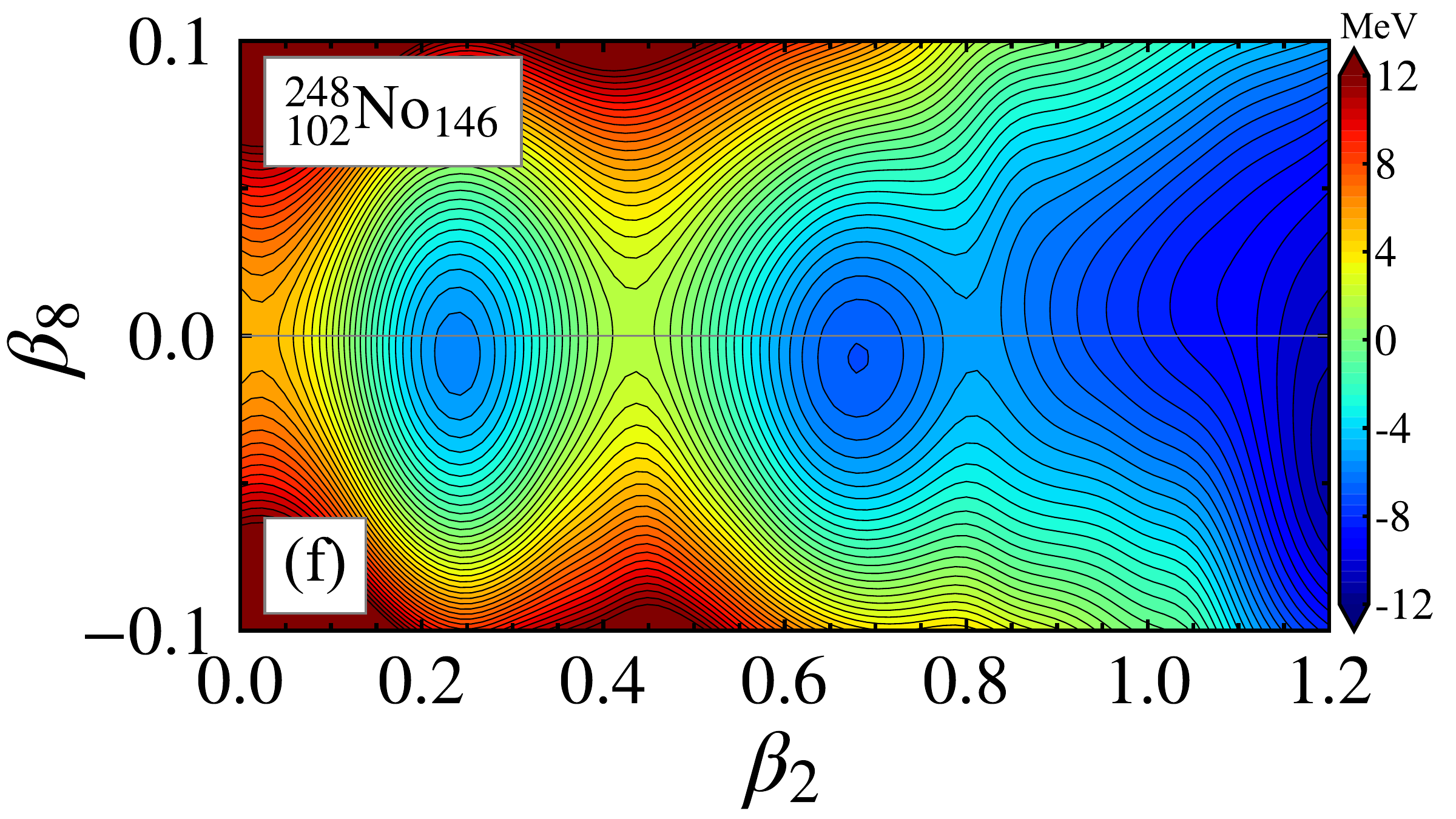}
	\caption{Similar to Fig.~\ref{Fig.3}, but for the $^{248}$No nucleus.}
	\label{Fig.4}
\end{figure*}

Quantum shell effects arising from single-particle states can enhance nuclear stability in the MM calculation because high and how densities will respectively give rise to positive and negative shell corrections. The appearance of these appropriate corrections on the pathway to scission may lead to enhanced stability. Figure~\ref{Fig.2} shows the effects of different high-order deformations (e.g., $\lambda \geq 5$) on micrscopic single-particle energies (not far from the Fermi surface) for neutrons (similarly, for protons) in the example nucleus $^{248}$No, indicating the success of the modified PES approach. The neutron spherical shell gaps at $Z$ = 126, 164 and 184 are reproduced from the single-particle energy diagram. The single-particle levels as a function of low-order deformations (e.g., $\beta_2, \gamma$ and $\beta_4$) can be easily found in the literature, i.e., cf. Refs~\citep{Chai2018a}. It deserves noticing that the nucleus with odd $\lambda$ deformation possesses the same shapes {(namely, the same nuclear potential but different orientation)} for positive and negative $\beta_\lambda$ values; the Hamiltonian of nuclear system will satisfy the relation $\hat{H}(-\beta_{\lambda_{odd}})$ = $\hat{H}(+\beta_{\lambda_{odd}})$ so that the single-particle diagram is symmetric about positive and negative $\beta_{\lambda_{odd}}$ values. Of course, it can be easily imagine that, for the separate deformation $\beta_{\lambda_{odd}}$, the macroscopic liquid-drop energy $E_{ld}$ also satisfies $E_{ld} (+\beta_{\lambda_{odd}}) = E_{ld}(-\beta_{\lambda_{odd}})$ since the atomic nucleus at this moment just has the different spatial orientations. It may be somewhat complex for the combination of several odd-$\lambda$ deformations but one can determine the symmetry relations. For instance, in the three-dimensional subspace ($\beta_3, \beta_5, \beta_7$), it can be divided into eight sections (quadrants). The potential energies at the lattices  (+$\beta_3, +\beta_5, +\beta_7$) and  (-$\beta_3, -\beta_5, -\beta_7$) will be equal since the nuclear shapes are identical at these two kinds of deformation grids, abbreviated to ($+,+,+$) and ($-,-,-$) for short. Similarly, other three pairs of symmetric combinations will be ($+,+,-$) and ($-,-,+$), ($+,-,-$) and ($-,+,+$), ($-,+,-$) and ($+,-,+$). Such a symmetry will reduce the number of calculated deformation grids by half.

All projected two-dimensional $\beta_2$-vs-$\beta_\lambda$ ($\lambda=3,4,5,6,7$ or 8) maps for $^{246}$No and $^{248}$No in such a seven-dimensional deformation space have respectively been illustrated in Figs.~\ref{Fig.3} and \ref{Fig.4}. In each subplot, the total energy is minimized over the remaining deformation degrees of freedom (e.g., on the $\beta_2$-vs-$\beta_3$ plane, the energy is minimized over $\beta_{4,5,6,7,8}$). From these projection maps, some properties, e.g., energy minima and fission paths, can be analyzed. It should be noted that, ignoring the interpolated errors, the corresponding minima, e.g., the normally-deformed minina near $\beta_2=0.2$ and the superdeformed minima  near $\beta_2= 0.7$, in different projection maps are same. { In this two figures, all the odd-order deformation parameters are zero both at the normally-deformed and superdeformed minima. The ensembles ($\beta_2$, $\beta_4$, $\beta_6$, $\beta_8$; $E_\mathrm{min}$) are respectivly (0.234, 0.041, -0.017, -0.005; -5.82 MeV) and (0.679, 0.045, -0.005, -0.007; -7.08 MeV) for the normally-deformed and superdeformed minima in Fig.~\ref{Fig.3} [similarly, (0.242, 0.033, -0.021, -0.007; -6.33 MeV) and (0.679, 0.037, -0.003, -0.007; -7.12 MeV) in Fig.~\ref{Fig.4}].}  Since there is no experimental deformation information for these two nuclei, it is instructive to confront our calculations (or part of them) with other theory. Indeed, the equilibrium deformations of the normally deformed minima calculated by us are in good agreement with the results given by Moller et al.~\cite{Moller1995}. In Ref.~\cite{Moller1995}, nuclear ground-state deformations are calculated in the deformation space \{$\beta_\lambda; \lambda=2,3,4,6$\} based on the finite-range droplet macroscopic model and folded-Yukawa single-particle microscopic model; the calculated ($\beta_2, \beta_3, \beta_4, \beta_6)$ values are
(0.224, 0.00, 0.054, -0.025) and (0.235, 0.00, 0.048, -0.033), respectively, agreeing with the present results, as seen in Figs.~\ref{Fig.3} and \ref{Fig.4}. Further, our calculations also indicate that, besides $\beta_4$ and $\beta_6$, even-order deformation $\beta_8$ still has a slight impact on the normally deformed minima in these two nuclei. In addition, one can see that all the even-order deformations will affect the superdeformed energy minima to some extent. In particular, it seems that the impact of high-order deformation $\beta_8$ is more important then $\beta_6$, agreeing with the case illustrated in Fig.~\ref{Fig.1}(b) (in the strongly elongated situation, the nucleus may be softer along $\beta_8$ than $\beta_6$). Concerning the odd-order deformations $\beta_3, \beta_5$ and $\beta_7$, we find that they do not affect both the normally deformed and the superdeformed minima but affect the saddle-point positions and fission paths after the superdeformed minima.   

\begin{figure}[htbp]
	\centering
	\includegraphics[width=0.85\linewidth]{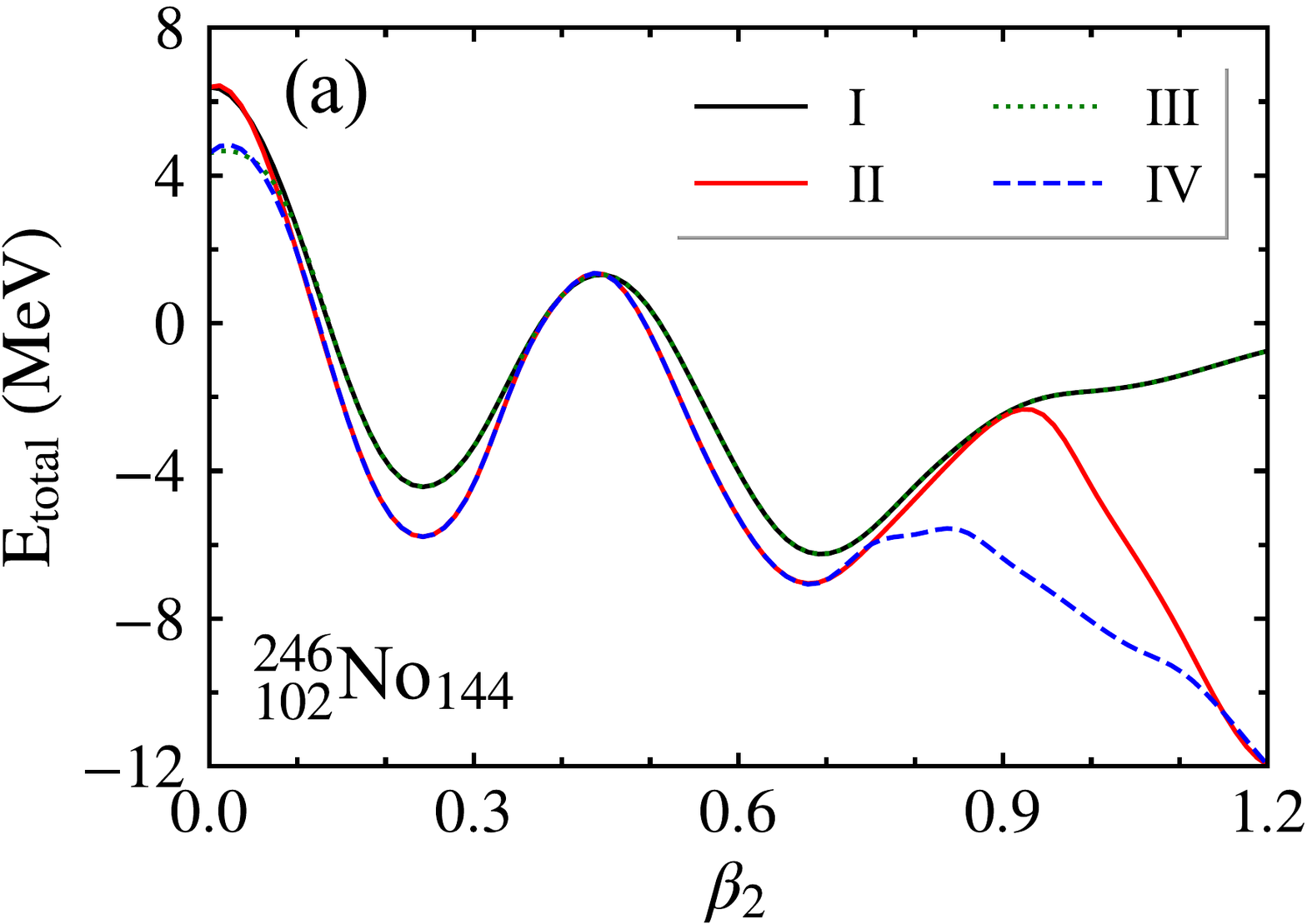}\\
	\vspace{-0.8cm}
	\includegraphics[width=0.85\linewidth]{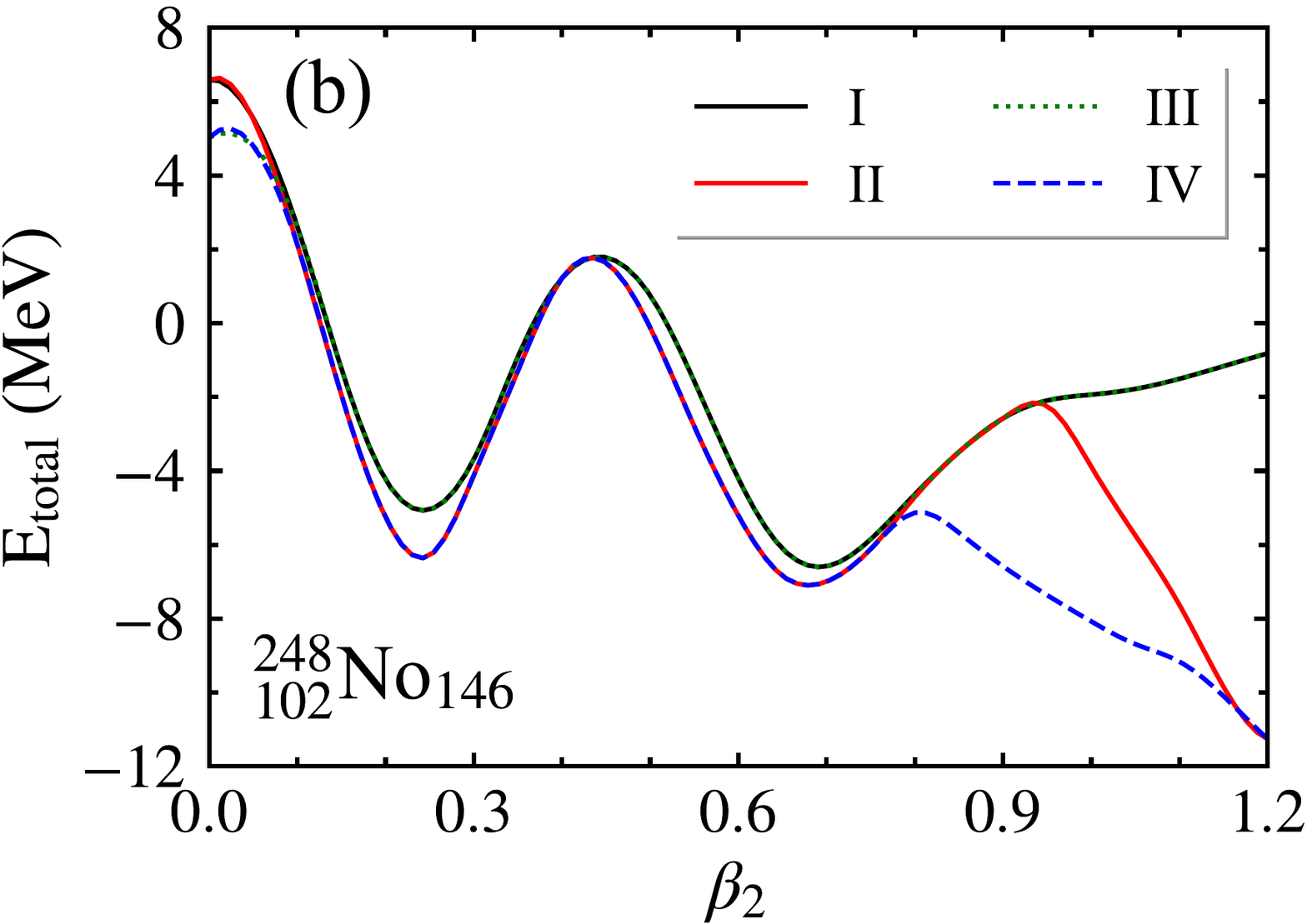}
	\caption{Four types of potential-energy curves as the function of deformation $\beta_2$ for $^{246}$No (a) and $^{248}$No (b). At each $\beta_2$ point, the energy minimization was performed over \{none\} (I; solid black line), $\{\beta_\lambda;\lambda=4,6,8\}$ (II; solid red line), $\{\beta_\lambda;\lambda=3,5,7\}$ (III; dotted green line) and $\{\beta_\lambda;\lambda=3,4,5,6,7,8\}$ (IV; dash blue line). See text for
further explanations.
}
	\label{Fig.5}
\end{figure}
\begin{figure}[htp]
	\centering
	\includegraphics[width=0.85\linewidth]{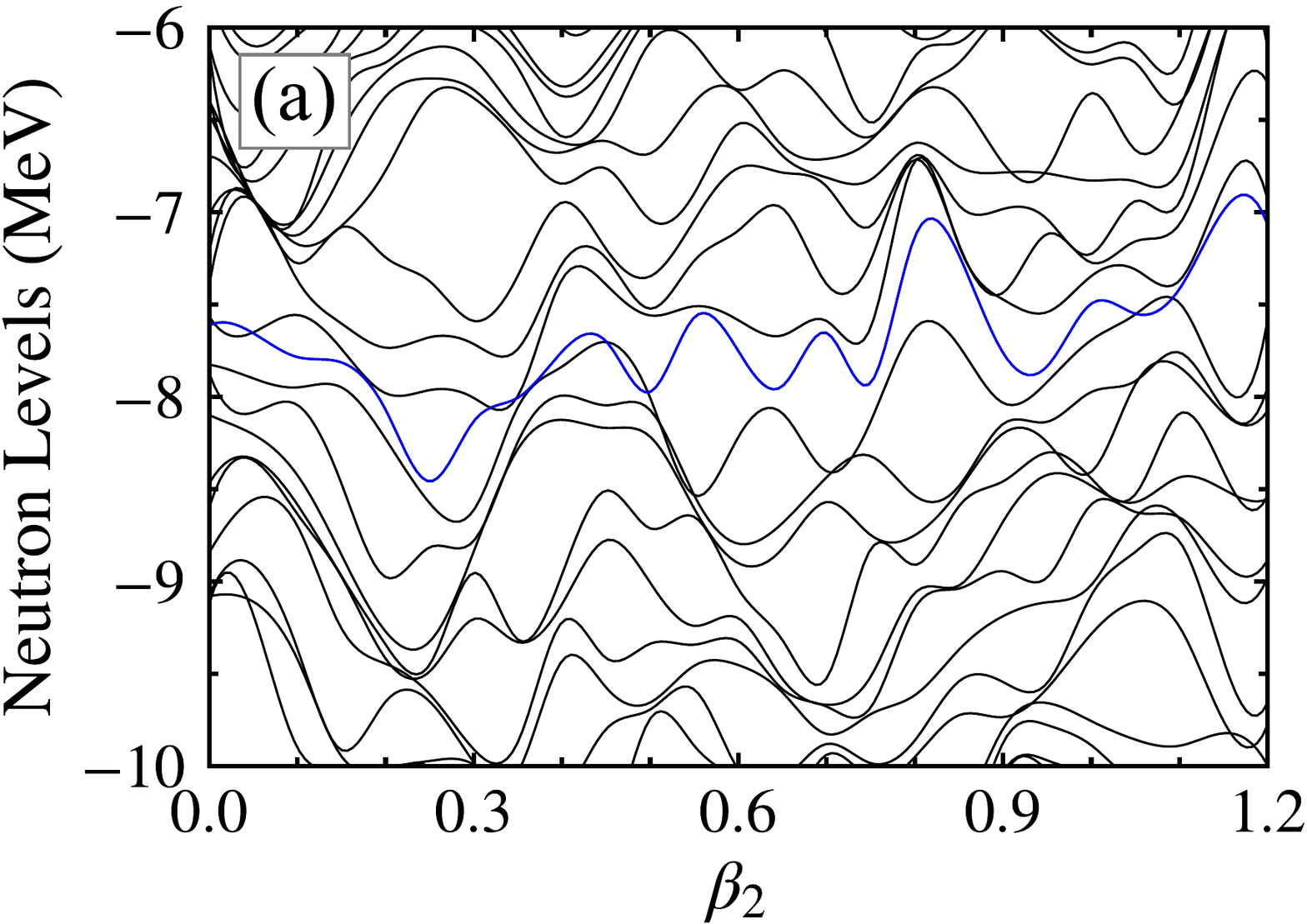}\\
	\vspace{-0.80cm}
	\includegraphics[width=0.85\linewidth]{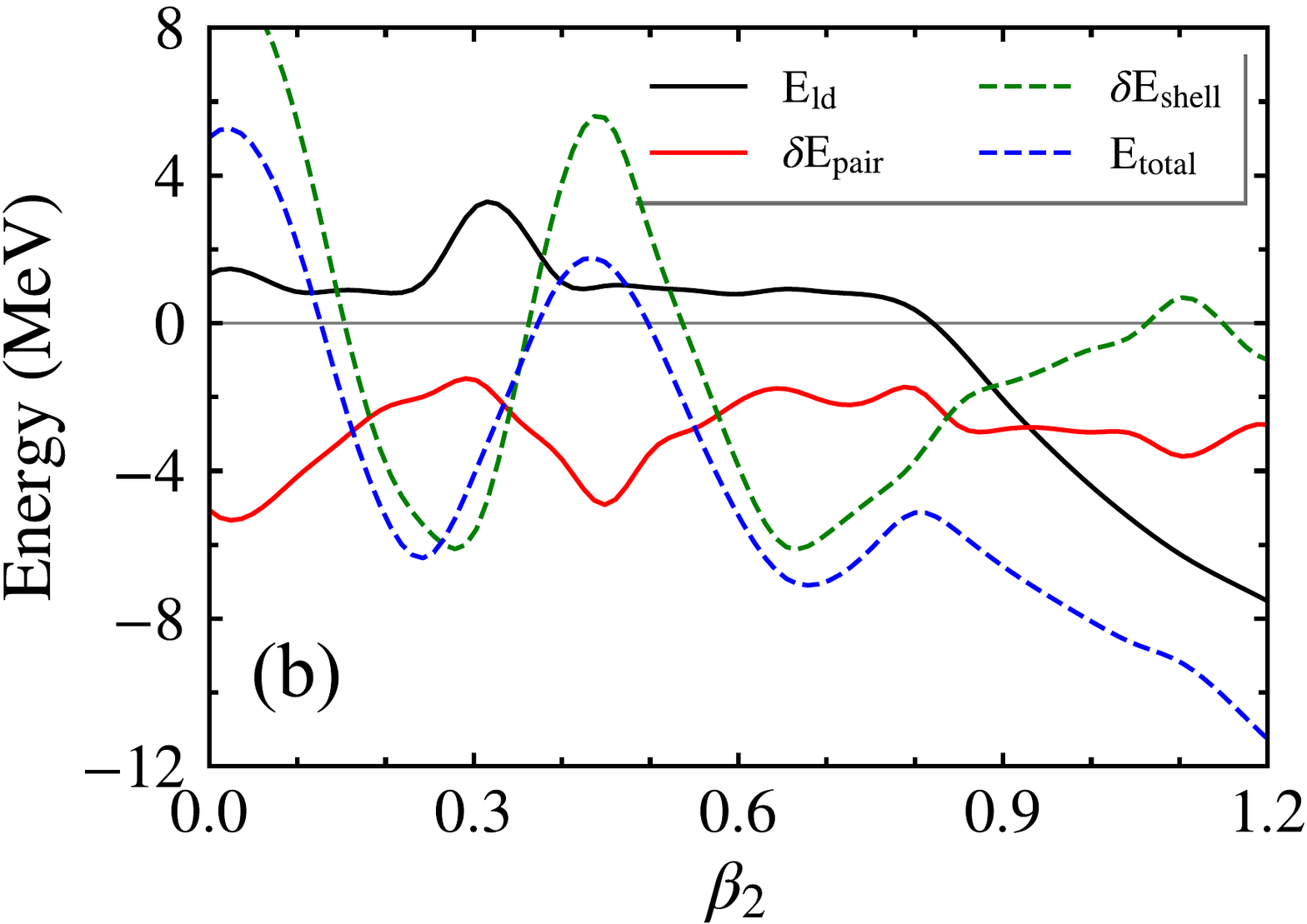}
	\caption{Neutron single-particle levels (a) and different energy curves (total energy and its macroscopic and microscopic compoments) in functions of $\beta_2$ for the nucleus $^{248}$No. Note that both for (a) and (b), at each $\beta_2$ grid, other deformation parameters adopt the values after the energy minimization and the total energy curve in (b) is same to the curve IV in Fig.5. In subplot (a), the energy level with blue color denotes the Fermi level.}
	\label{Fig.6}
\end{figure}

To understand the effects of even- and odd-order deformations on the fission trajectory, Figure~\ref{Fig.5} shows four types of potential-energy curves along the minimum valley in the quadrupole deformation $\beta_2$ direction for $^{246,248}$No. The typically double-humped fission barriers in actinide nuclei are well reproduced~\cite{Bjornholm1980}. Note that the $E(\beta_2)$ curves I and IV will respectively occupy the highest and lowest positions at each $\beta_2$ point since the former minimizes over \{none\} but the latter over \{$\beta_\lambda$; $\lambda=3,4,5,6,7,8$\}. Keeping this in mind, one can easily read this figure though there is a strong overlap of different curves. It can be seen that curve II, which is minimized over the remaining even-order deformations $\beta_{4,6,8}$, will further decrease the energies of not only the normally deformed minimum but also the superdeformed minimum and leads to the formation of the second barrier. Indeed, only considering the deformation $\beta_2$, e.g., curve I, the energy will continue increasing after the superdeformed minimum with the increase of $\beta_2$ (at least, up to $\beta_2=1.2$). Except for the weak deformation region (e.g., approximately $\beta_2 \leq 0.1$), the energy curves I and III fully overlap, indicating that there is no odd-order deformation effects along the fission valley in the deformation subspace ($\beta_2, \beta_3, \beta_5, \beta_7$). Similarly, the overlap of the energy curves II and IV before $\beta_2\approx 0.8$ also illustrate such negligible odd-order deformation effects. Comparing curves II and IV, one can find that the inclusion of odd-order deformations will further decrease the second barrier, indicating the occurrence of the coupling between odd- and even-order deformations. Obviously, the properties of the potential-energy curves are similar for subplots (a) $^{246}$No and (b) $^{248}$No, except for a slightly lower outer-barrier in $^{246}$No. As known, both spontaneous fission and $\alpha$ decay, terminating the stability of drip-line heavy nuclei, sensitively depend on such potential-energy curves. With the decrease of neutron number, $^{246}$No is excepted to have a shorter half-life than $^{248}$No but there should be no a abrupt reduction according to their properties of fission trajectories. {Certainly, it is instructive to investigate the evolution properties of single-particle energies, macroscopic and microscopic energies along the ``realistic'' fission path (corresponding to the curve IV in Fig.~\ref{Fig.5}). Accordingly, the representative single-neutron diagram and different energy curves are illustrated for $^{248}$No (similarly for $^{246}$No) in Fig.~\ref{Fig.6}. One can find that the single-particle levels, involving different deformations, become more complicated, as seen in Fig.~\ref{Fig.6}(a). From Fig.~\ref{Fig.6}(b), it can be seen that the formation of the inner barrier primarily originates from the microscopic shell correction and the outer barrier is strongly affected by macroscopic liquid-drop energy and microscopic shell correction. The pairing correlation always provides a negative and relatively-smoothed energy.}

\begin{figure}[ht]
\centering
\includegraphics[width=1.0\linewidth]{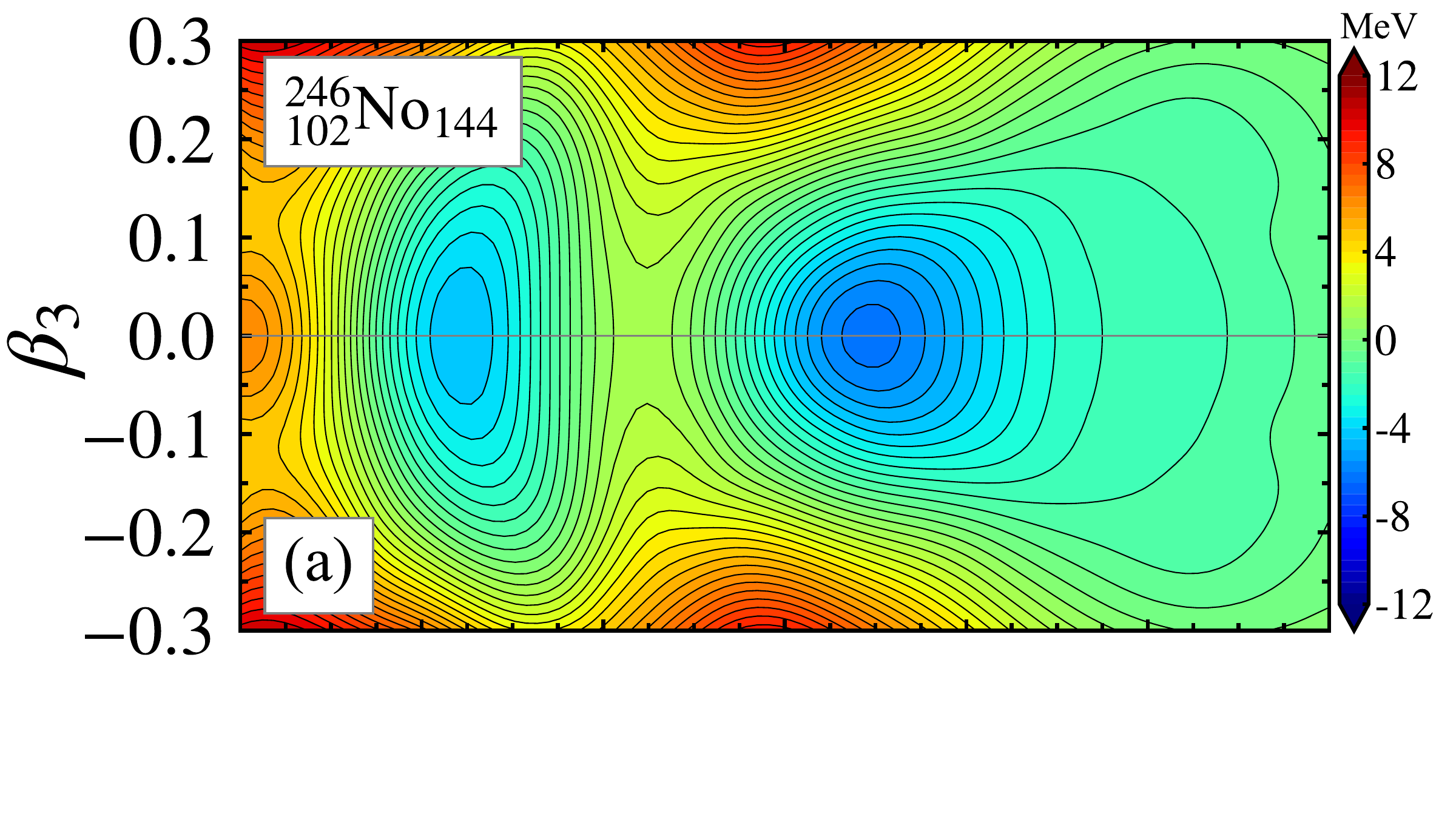}\\
\vspace{-1.0cm}
\includegraphics[width=1.0\linewidth]{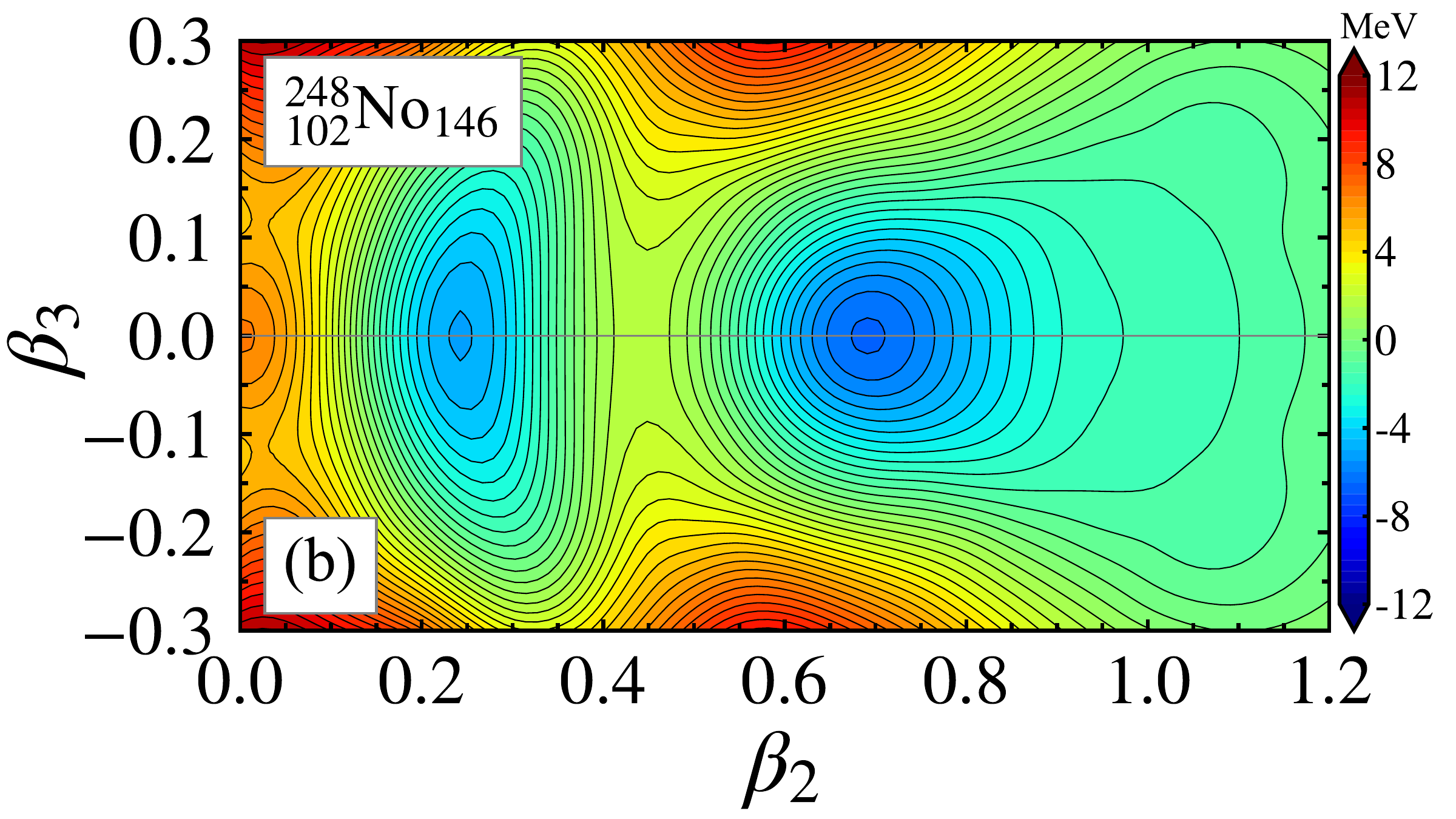}
\caption{Similar to Fig.~\ref{Fig.3}(a) and Fig.~\ref{Fig.4}(a), potential-energy projections on ($\beta_2$, $\beta_3$) plane for $^{246}$No (a) and $^{248}$No (b). But, in each subplot, the minimization is performed over $\beta_5$ and $\beta_7$, without the consideration of even-order deformations.}
	\label{Fig.7}
\end{figure}

It should be pointed out that though the normally deformed minimum in $^{246,248}$No is still referred to as ground state in Ref~\cite{Moller1995}, the inversion of the energies between the normally deformed minimum and the superdeformed minimum has occurred. Therefore, strictly speaking, the superdeformed minima in $^{246,248}$No are their ground states. Actually, the fission half-lives of $^{246,248}$No decayed from such ground states will rapidly decrease, relative to from the normally-deformed ground states (e.g., in the lighter actinide nuclei). However, as discussed in Ref~\cite{Xu2004} where it is pointed out that the stability of superheavy nuclei may be enhanced by high-K isomer, such very neutron-deficient heavy nuclei may have the enhanced stability due to the normally-deformed minimum as the shape isomer (The study of decay half-life from it, including the typical $\gamma$ distortion of the inner barrier, e.g., cf. Ref.~\cite{Chai2018a}, is beyond the scope of the present work).

In addition, in order to verify that odd-order deformation effects for the fission paths can occur only when the even-order deformations are considered, we show the energy projection maps in the ($\beta_2$, $\beta_3$) plane for $^{246,248}$No in Fig.~\ref{Fig.7}, ignoring the even-order deformation degrees of freedom. The fission valley in Fig.~\ref{Fig.7} is equivalent to the curve III in Fig.~\ref{Fig.5}. From Fig.~\ref{Fig.7}, one can see that, from the normally deformed minima to the strongly elonged region, the odd-order deformation $\beta_3$ does not change the fission path in $^{246,248}$No. It can be concluded that the odd-order deformation effects only play important roles, accompanying the higher even-order deformations (e.g., $\beta_4$). {Whether such a conclusion is a general rule deserves further study by a systematic investigation in the future.}     

\section{Summary}
\label{summary}

In this project, we have developed the PES calculation method, extending the deformation space, within the framework of the MM model and investigated the high-order deformation effects in the neutron-deficient heavy nuclei $^{246,248}$No. The evolution properties of microscopic single-particle levels and macroscopic energies in functions of different deformation degrees of freedom are illustrated. It is found that the higher the deformation order is, the more difficult it is to occur since, for a spherical liquid drop, in general, the stiffness along some deformation will gradually increase as the corresponding deformation-multipolarity increases. However, for a strongly elongated spheroid, the high-order deformations will play an important role owing to the large softness along them. Our calculations illustrate that the highly even-order deformations significently affect both the potential energy minima and fission paths. In particular, the high-order deformation $\beta_8$ may be more favored than the lower-order $\beta_6$ at the strongly elongated nuclear shape. All the odd-order deformations will mainly have an impact on the second barrier but they must accompany the even-order ones (e.g., $\beta_4$). Indeed, the inclusion of high-order deformations is somewhat necessary in the study of nuclear structure and nuclear fusion and fission processes. { Though we cannot accurately determine the symmetry properties of fission fragments due to scarce information of the scission points, the trend in the earlier studies~\cite{Wild1994,Albertsson2020} indicates the possible occurance of asymmetric fissions in these two nuclei.} { Moreover, it is also found that the very neutron-deficient $^{246}$No nucleus may still be accessed experimentally since, similar to the high-K isomer reported in Ref.~\cite{Xu2004}, the normally-deformed shape isomer can enhance the survival probality in the drip-line heavy nucleus though the superdeformed ground state are rather unstable.} Of course, it is meaningful to further extend the deformation space including the nonaxially deformations in the future. \\

\section*{Conflict of Interest}

The authors declare that t{hey have no known competing financial
interests or personal relationships that could have appeared to
influence the work reported in this paper.
\label{key}

\section*{Acknowledgement}
This work was supported by the Natural Science Foundation of Henan Province (No. 252300421478) and the National Natural Science Foundation of China (No.11975209, No. U2032211, and No. 12075287). Some of the calculations were conducted at the National Supercomputing Center in Zhengzhou.
%

\end{document}